\documentclass[12pt,a4paper]{article}
\usepackage[latin1]{inputenc}
\usepackage{amsmath,bm}
\usepackage{amsmath} 
\usepackage{amsfonts}
\usepackage{amssymb}
\usepackage{float}
\usepackage{cite}
\usepackage{breqn}

\makeatletter
\newcommand{\mathleft}{\@fleqntrue\@mathmargin0pt}
\newcommand{\mathcenter}{\@fleqnfalse}
\makeatother

\usepackage[colorlinks=true,linkcolor=blue,citecolor=red]{hyperref}

\usepackage{graphicx}
\DeclareGraphicsExtensions{.pdf,.png,.jpg}

\usepackage[a4paper, width = 160mm, top = 30mm, bottom = 25mm, 
bindingoffset = 0mm]{geometry}

\def\beq{\begin{equation}}
\def\eeq{\end{equation}}
\def\bea{\begin{eqnarray}}
\def\eea{\end{eqnarray}}

\begin{document}

\begin{center}
  {\Large \bf Quantum tunneling from a new type of Unified Cantor Potential}
\vspace{1.3cm}

{\sf Mohammad Umar\footnote[1]{e-mail address:\ \ aliphysics110@gmail.com, opz238433@opc.iitd.ac.in\hspace{0.05cm}},
Vibhav Narayan Singh\footnote[2]{e-mail address:\ \ vibhav.ecc123@gmail.com, vibhavn.singh13@bhu.ac.in },
Mohammad Hasan\footnote[3]{e-mail address:\ \ mhasan@isro.gov.in,\hspace{0.05cm} mohammadhasan786@gmail.com},\\
Bhabani Prasad Mandal\footnote[4]{e-mail address:\ \ bhabani.mandal@gmail.com,\hspace{0.05cm} bhabani@bhu.ac.in}}

\bigskip

{\em $^{1}$Optics and Photonics Centre, Indian Institute of Technology, Delhi-110016, INDIA \\ 
$^{2,4}$Department of Physics,
Banaras Hindu University,
Varanasi-221005, INDIA. \\
$^{3}$Indian Space Research Organisation,
Bangalore-560094, INDIA. \\}

\bigskip	
	\noindent {\bf Abstract}
\end{center}
We introduce a new type of potential system that combines the families of general Cantor (fractal system) and general Smith-Volterra-Cantor (non-fractal system) potentials. We call this system as Unified Cantor Potential (UCP) system. The UCP system of total span $L$ is characterized by scaling parameter $\rho >1$, stage $G$ and two real numbers $\alpha$ and $\beta$. For $\alpha=1$, $\beta=0$, the UCP system represents general Cantor potential while for $\alpha=0$, $\beta=1$, this system represent general Smith-Volterra-Cantor (SVC) potential. We provide close-form expression of transmission probability from UCP system for arbitrary $\alpha$ and $\beta$ by using $q$-Pochhammer symbol. Several new features of scattering are reported for this system. The transmission probability $T_{G}(k)$ shows a scaling behavior with $k$ which is derived analytically for this potential. The proposed system also opens up the possibility for further generalization of new potential systems that encompass a large class of fractal and non-fractal systems. The analytical formulation of tunneling from this system would help to study the transmission feature at breaking threshold when a system transit from fractal to non-fractal domain.  

\newpage
\section{Introduction}
Quantum tunneling, a cornerstone of early quantum mechanics research, has captivated researchers since its inception in 1928 \cite{nordheim, gurney} and has remained a fundamental topic in the field. Over the course of time, numerous researchers have made significant contributions to our understanding of quantum tunneling, enriching the knowledge of its underlying principles \cite{condon, wigner, bohm, book1, book2}. Extensive investigations have been conducted to explore the propagation of matter waves through various potential distributions, driven by the theoretical significance, experimental relevance, and practical applications associated with this phenomenon. As a result, a diverse range of potentials have been examined, and different methods have been developed to address a wide spectrum of tunneling problems \cite{book1, book2}. Different authors for over a century have contributed to various aspects of quantum tunneling \cite{nordheim, gurney, condon, wigner, bohm, esaki, giaever, burstein, josephson, book1, lauhon, book2}. Tunneling features of a quantum particle through a given distribution of potential by the analytical calculation of scattering coefficients have been explored in other domains of quantum mechanics such as non-Hermitian quantum mechanics (NHQM) \cite{angelopoulou1995non,hasan2020hartman,hasan2020role,longhi2022non}, space-fractional quantum mechanics (SFQM) \cite{guo, oli, tare,hasan2020tunneling1,hasan2020tunneling2} and quaternionic quantum mechanics (QQM) \cite{sobhani,hasan2020new, hassan01, sobhani01, de, davis, de01}. The analytical calculations of the transmission coefficient for a potential system provide a deeper understanding of transmission features as compared to the numerical methods. Transmission of  quantum waves through fractal potentials shows many interesting features. Some of the important features are sharp transmission peaks and self-similarity in the transmission profile with wave vector  \cite{feder, wen, shalaev1, shalaev2, shalaev3, takeda, chuprikov2000, miyamoto, cantor_graphene, voss, hurd}. Due to experimental observations of theoretical features of tunneling from fractal systems \cite{takeda, miyamoto}, quantum tunneling through fractal potentials has gained much attention.\\
\indent
Geometrical objects which preserve its self-similarity and homogeneity feature in its sub-components are called fractals. The term fractal is first coined by mathematician Benoit B. Mandelbrot \cite{mandelbrot}. Performing the basic mathematical operation recursively on a given geometrical object leads to the construction of self-similar structures or fractals. In this context, the mathematical operation is known as \textit{generator} and the geometrical object is known as \textit{initiator}. Mathematical operation is performed on multiple levels and as a result, each level is recognized by the sub-components of the original object (\textit{initiator}) having a complete resemblance of the whole original object showing the feature of self-similarity. Self-similar objects are scale-invariant and this property holds at all scales for mathematical objects.  Occurrences of fractals are also seen in nature and the structure of many naturally occurring objects are approximated with the use of fractals.  \cite{mandelbrot, voss, hurd}.\\
\indent
Cantor fractal potential is a member of the fractal family and is the simplest fractal potential. Starting from a rectangular potential of height $V$ and length $L$ (this is stage $G=0$),  one can obtain a stage $G$ general Cantor (GC) fractal potential by consecutively removing $\frac{1}{\rho}$ fraction ($ \rho \in \mathbb{R}^{+}$ and $\rho >1$) of length segment from the middle at all steps $G=1,2,3,...,G$ from the remaining segments of the potential. When $\rho = 3$, it represents the \textit{standard} Cantor fractal potential simply known as Cantor fractal potential or  Cantor potential. There is another potential distribution analogous to Cantor fractal potential named as general Smith-Volterra-Cantor (GSVC or SVC-$\rho$) potential where a fraction $\frac{1}{\rho^{G}}$ of the length of the potential segment of stage $G-1$ is removed from the middle of each remaining potential segments at every stage $G$. When $\rho =4$, we have the \textit{standard} SVC-$4$ potential system. SVC-$\rho$ potential system does not belong to the family of fractal system as it doesn't preserve the same structure of self-similarity at different stages. Using the concept of transfer matrix method in quantum mechanics, scattering properties through Cantor fractal potential has been studied by many authors \cite{cantor_f1, cantor_f2, cantor_f3, cantor_f4, cantor_f5, cantor_f6, cantor_f7, cantor_f8, cantor_f9}. The composition properties of the transfer matrix have been used to derive the scattering coefficients and associated properties. In Cantor fractal potential system, the appearance of scaling of the transmission coefficient with respect to the wave vector $k$ are seen \cite{cantor_f1, cantor_f6, cantor_f7}.\\
\indent
In earlier work, we introduced the concept of super periodic potential (SPP) \cite{mh_spp} which is the generalization of the locally periodic potential. Using this formalism one can easily obtain the close form expression of the reflection and transmission amplitude for an arbitrary order of periodicity of a potential  provided that the transfer matrix of the corresponding \textit{unit cell} potential is known. It is shown that the general Cantor and SVC-$\rho$  potential system are a special case of the SPP system.  By using the SPP formalism, the transmission coefficients from general Cantor and SVC-$\rho$ potential are studied both for the case of standard QM \cite{mh_spp, vn_svc} and general as well as for the case of SFQM \cite{vn_sfqm}.\\ 
\indent
In this article, we  introduce a more general class of Cantor potential system that unifies the  general Cantor and SVC-$\rho$ system. We name this potential as Unified Cantor Potential (UCP) and provide the transmission coefficient from UCP by using SPP formalism developed earlier. The construction of UCP system of stage $G$ is similar to the construction of general SVC system of stage $G$.  The UCP system of stage $G$ is characterized by  four parameters $\alpha$, $\beta$, $\rho$ and $G$. UCP-$\rho$ system is constructed by removing $\frac{1}{\rho^{\alpha + \beta G}}$ fraction of the length of the potential segment of stage $G-1$ from the middle of each remaining portion of the potential at every stage $G$. Again here, $\rho$ is a real positive number and $\alpha$ and $\beta$ also belongs to the real number ($\alpha$, $\beta$ $\in$ $\mathbb{R}$) with condition that $\alpha$ and $\beta$ can not be simultaneously zero. Zero value of $\alpha$ and $\beta$ removes the whole potential at stage $G = 1$ leaving behind an empty space. Here, $\alpha = 1$, $\beta = 0$ will represent general Cantor fractal system and $\alpha = 0$, $\beta = 1$ will SVC-$\rho$ potential system. Hence general Cantor and SVC-$\rho$ potential systems are the special cases of the UCP-$\rho$ system. In other words, the UCP-$\rho$ system combines a fractal and non-fractal system together and therefore it also presents an interesting system to study the transmission features from this system.\\    
\indent
We organize the paper as follows: In Section \ref{cantor}, we introduce the concept of the UCP-$\rho$ system in detail. Section \ref{ucpspp} explains how UCP-$\rho$ system of arbitrary stage $G$ is a special case of rectangular SPP of order $G$. In Section \ref{tucp}, explicit expression of $\Omega_{q}$ is expressed in order to get transmission probability for UCP-$\rho$ system. Additionally, we discussed general Cantor and SVC-$\rho$ potential system as a special case of UCP-$\rho$ system. In Section \ref{transfeatures} we provide graphically a detailed analysis of transmission features and associated results have been discussed in this section. Finally, the paper culminates with the conclusion and discussions presented in Section \ref{conclusion}.
\section{Unified Cantor potential system}
\label{cantor}
\begin{figure}[H]
\begin{center}
\includegraphics[scale=0.40]{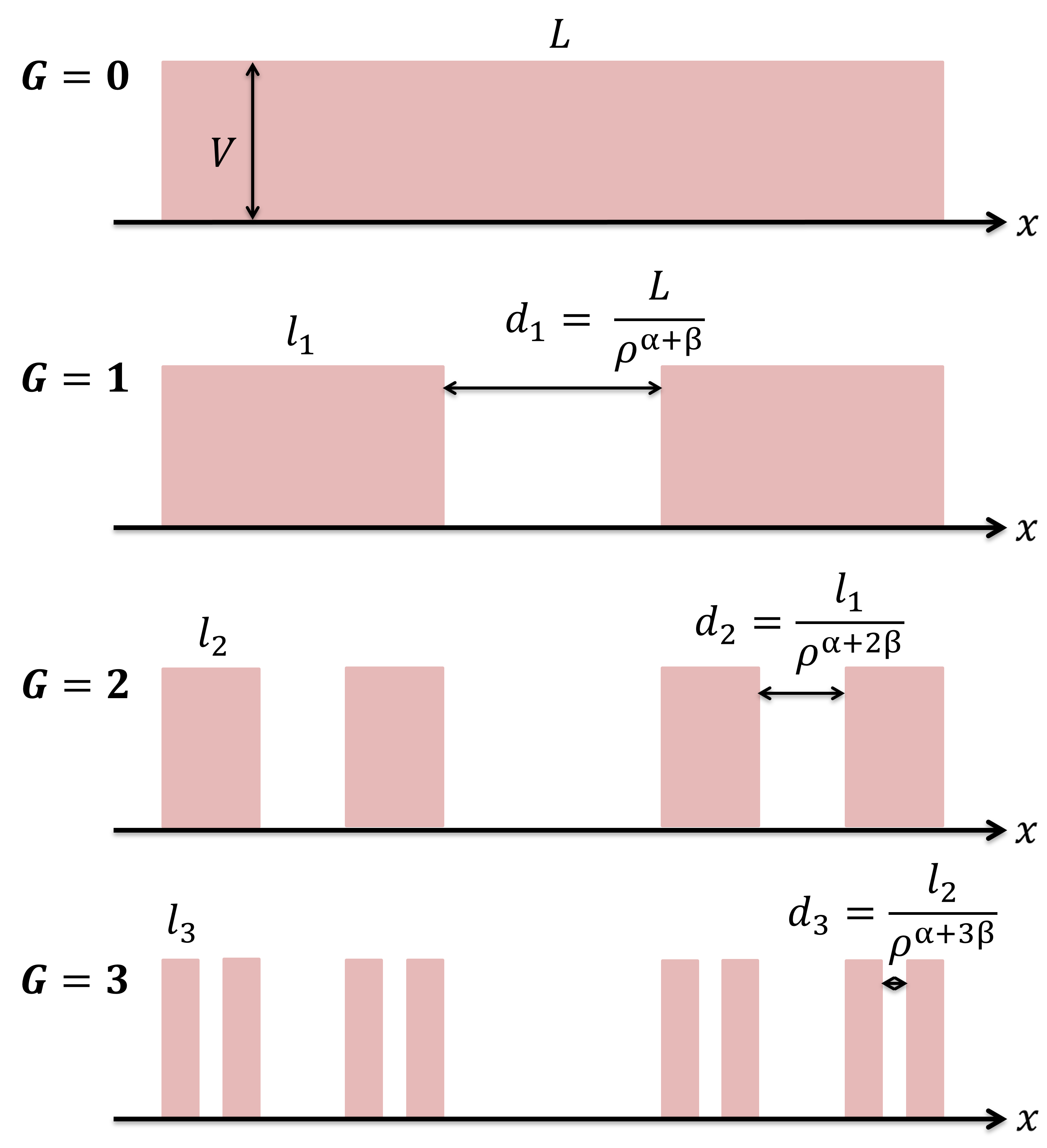} 
\caption{\it Construction of the UCP-$\rho$ system. The white region shows the gap between the potentials and the height of the opaque region is the potential height V. Here G represents the stage of the system. $l_{G}$ and $d_{G}$ represent the length of each potential segment and gap between the potentials at $G^{th}$ stage respectively.}
\label{general_figure}
\end{center}
\end{figure}

We introduce the concept of a unified potential system that joins the family of general Cantor potential (a fractal potential) and general SVC potential (non-fractal potential). We call this system as Unified Cantor Potential (UCP). To begin with, consider a rectangular barrier potential of finite length $L$ and height $V$. The proposed UCP-$\rho$ system is obtained by removing the fraction $\frac{1}{\rho^{\alpha + \beta G}}$ from the middle of each segment at each stage $G$. In this case, $\rho$ is a real positive number and greater than one. $\alpha$ and $\beta$ are real numbers, provided that they can not be simultaneously zero. The formation of this UCP-$\rho$ system is depicted in Fig. \ref{general_figure}. Stage $G = 1$ illustrates the removal of a fraction $\frac{1}{\rho^{\alpha + \beta }}$ from the middle of the potential of height $V$ and length $L$ as shown in Fig. \ref{general_figure}. Stage $G = 1$ has two potential segments of height $V$ and length $l_{1}$. For the next stage $G = 2$, further a fraction $\frac{1}{\rho^{\alpha + 2\beta }}$ is eliminated from the middle of the remaining two potential segments of stage $G = 1$ and in this case, each potential segment has length $l_{2}$. For stage $G = 3$, a fraction $\frac{1}{\rho^{\alpha + 3\beta }}$ is eliminated from the middle of the remaining four potential segments of the stage $G = 2$ and in this case, each potential segment has length $l_{3}$. In general, applying the same process of elimination of the fraction $\frac{1}{\rho^{\alpha + \beta G}}$ from the middle of the remaining each potential segments of stage $G-1$ constructs the UCP-$\rho$ system of any arbitrary stage $G$ consisting of $2^{G}$ potential segment of equal length $l_{G}$.
From Fig. \ref{general_figure}, it is clear that the length of the potential segment at stage $G=1$ is expressed through,
\begin{equation}
l_{1} = \frac{L}{2}\left(1-\frac{1}{\rho^{\alpha+\beta}}\right).
\end{equation}
Similarly, the length of the potential segment at stage $G=2$ is expressed through
\begin{equation}
l_{2} = \frac{L}{2^{2}}\left(1-\frac{1}{\rho^{\alpha+\beta}}\right)\left(1-\frac{1}{\rho^{\alpha+2\beta}}\right).
\end{equation}
Similarly, length of the potential segment at stage $G=3$ is expressed through
\begin{equation}
l_{3} = \frac{L}{2^{3}}\left(1-\frac{1}{\rho^{\alpha+\beta}}\right)\left(1-\frac{1}{\rho^{\alpha+2\beta}}\right)\left(1-\frac{1}{\rho^{\alpha+3\beta}}\right).
\end{equation}
and so on. 
Similarly, length of the potential segments at stage $G=4$ is expressed through
\begin{equation}
l_{4} = \frac{L}{2^{4}}\left(1-\frac{1}{\rho^{\alpha+\beta}}\right)\left(1-\frac{1}{\rho^{\alpha+2\beta}}\right)\left(1-\frac{1}{\rho^{\alpha+3\beta}}\right)\left(1-\frac{1}{\rho^{\alpha+4\beta}}\right).
\end{equation}
In general, length $l_{G}$ of potential segment of any arbitrary stage $G$ can be expressed as
\begin{equation}
l_{G} = \frac{L}{2^{G}}\prod_{j=1}^{G}\left(1-\frac{1}{\rho^{\alpha+\beta j}}\right).
\label{general_lG}
\end{equation}
The product series can be expressed in terms of $q$-Pochhammer symbol as
\begin{equation}
\prod_{j=1}^{G}\left(1-\frac{1}{\rho^{\alpha+\beta j}}\right) = \frac{\rho^{\alpha}}{\rho^{\alpha}-1}\times q\left(\frac{1}{\rho^{\alpha}};\frac{1}{\rho^{\beta}}\right)_{G+1},
\end{equation}
where $q$-Pochhammer symbol is defined by,
\begin{equation}
q(\mu;\nu)_{p}=\prod_{j=0}^{p-1}(1-\mu.\nu^{j})=(1-\mu)(1-\mu.\nu)(1- \mu.\nu^{2}).....(1-\mu.\nu^{p-1}).
\label{qp}
\end{equation}
Hence $l_{G}$ in terms of $q$-Pochhammer symbol is expressed through
\begin{equation}
l_{G} = \frac{L\rho^{\alpha}}{2^{G}(\rho^{\alpha}-1)}\times q\left(\frac{1}{\rho^{\alpha}};\frac{1}{\rho^{\beta}}\right)_{G+1}.
\label{general_lg}
\end{equation}
\section{Unified Cantor potential (UCP-$\rho$) as a special case of super periodic rectangular potential}
\label{ucpspp}
The concept of super periodic potential (SPP) is introduced in \cite{mh_spp}. However, for the sake of completeness, we briefly mention the concept in this paper. Starting from a unit cell potential $V$, a periodic system is obtained by the periodic repetitions of the unit cell system to a finite number $N_{1}$ times at consecutive distances $s_{1}$. Here $s_{1}$ is the distance between the starting point of the two consecutive unit cells. The periodic potential is denoted as $V_{1}=(V, N_{1},s_{1})$. Now the system $V_{1}$ is considered as a unit cell and periodically repeated with number $(N_{2}, s_{2})$ to obtain the system $V_{2}=(V_{1}, N_{2},s_{2})$ which is further periodically repeated to obtain the system $V_{3}=(V_{2}, N_{3}, s_{3})$. The process of periodic repetitions can continue to an arbitrary finite number of times $g$ to obtain the SPP system of order $g$, $V_{g}=(V_{g-1}, N_{g}, s_{g})$. 

\begin{figure}[H]
\begin{center}
\includegraphics[scale=0.47]{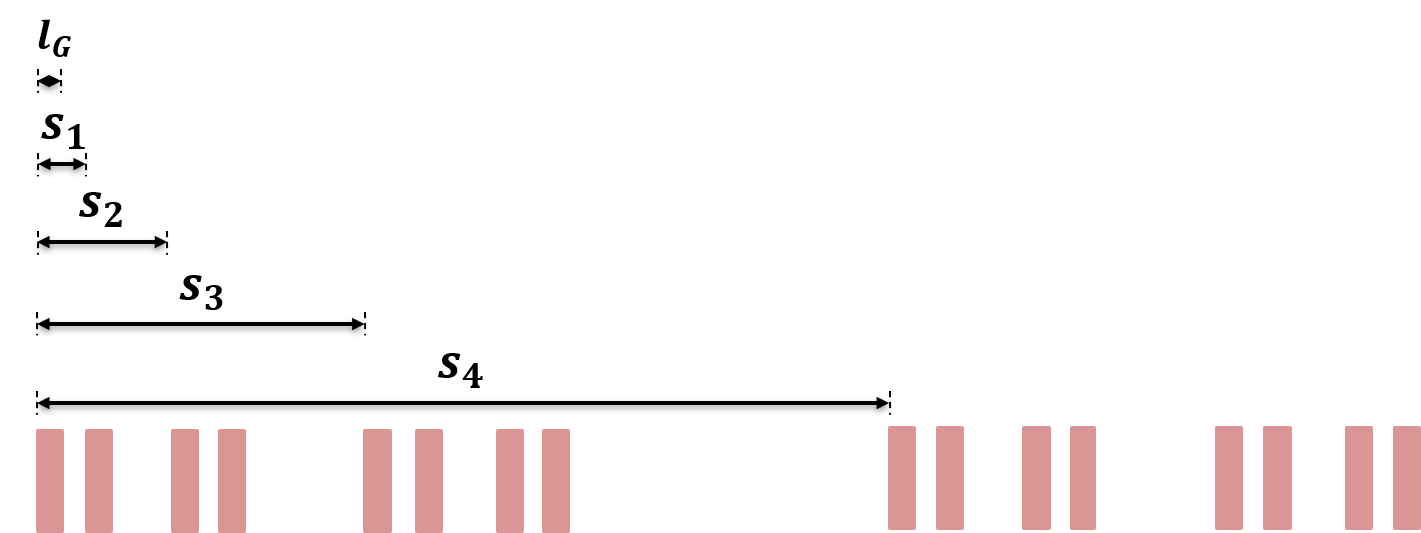} 
\caption{\it Construction of the UCP-$\rho$ system of stage $G=4$ through super periodic manner. Here initiator is a rectangular barrier of height $V$ and the generator is super periodicity. $s_{1}$, $s_{2}$, $s_{3}$ and $s_{4}$ is the super periodic distance and the number of periodic repetitions is 2.}
\label{general_figure1}
\end{center}
\end{figure}

Next, we show that the UCP-$\rho$ system of stage $G$ is the special case of rectangular SPP of order $G$. As discussed earlier, the UCP-$\rho$ system 
is obtained by removing the fraction $\frac{1}{\rho^{\alpha + \beta G}}$ from the middle of each segment at each stage $G$. There are two potentials of equal spatial length $l_{1}$ at a separation of $d_{1}$ distance, which implies a periodic potential system with $N=2$ and periodic distance $s_{1}=l_{1}+d_{1}$ (Fig. \ref{general_figure1}). As the construction of the UCP-$\rho$ system is based on the removal of the middle portion (of a specific length) from the potential segments progressively at each stage $G$ in such a fashion that each stage has the potential of the same length. This implies that the UCP-$\rho$ system is a special case of SPP with $N_{f}=2$ and with some $s_{f}$, $f=1,2,3,....,g$. Starting from a unit cell rectangular barrier, the construction of the UCP-$\rho$ system is shown in Fig. \ref{general_figure1} for stage $G=4$. This illustration shows that the total number of super periodic repetition $g$ is equal to the stage $G$ of the UCP-$\rho$ system and therefore $G=g$. Below we arrive at the calculation of $s_{f}$ as follows:
  
From the definition of UCP-$\rho$ system and from the geometry of Fig. \ref{general_figure} and \ref{general_figure1}, it is noted that,
\begin{equation}
s_{1} = l_{G} + \frac{l_{G-1}}{\rho^{\alpha+ \beta G}},
\end{equation}
\begin{equation}
s_{2} = l_{G-1} + \frac{l_{G-2}}{\rho^{\alpha+ \beta (G-1)}},
\end{equation}
\begin{equation}
s_{3} = l_{G-2} + \frac{l_{G-3}}{\rho^{\alpha+ \beta (G-2)}},
\end{equation}
\begin{equation}
s_{4} = l_{G-3} + \frac{l_{G-4}}{\rho^{\alpha+ \beta (G-3)}}.
\end{equation}
In general,
\begin{equation}
s_{f} = l_{G+1-f} + \frac{l_{G-f}}{\rho^{\alpha+ \beta (G+1-f)}}.
\end{equation}
Using Eq. (\ref{general_lG}) in above, we arrive at
\begin{equation}
s_{f} = \frac{L}{2^{G+1-f}}\prod_{j=1}^{G+1-f}\left(1-\frac{1}{\rho^{\alpha+\beta j}}\right)+\frac{L}{2^{G-f}\rho^{\alpha+\beta(G+1-f)}}\prod_{j=1}^{G-f}\left(1-\frac{1}{\rho^{\alpha+\beta j}}\right).
\end{equation}
After simplification, this can be expressed as,
\begin{equation}
s_{f} = \frac{L}{2^{G+1-f}}\left( 1+\frac{1}{\rho^{\alpha+\beta(G+1-f)}}\right)\prod_{j=1}^{G-f}\left(1-\frac{1}{\rho^{\alpha+\beta j}}\right).
\label{s_m}
\end{equation}
Further in terms of $q$-Pochhammer symbol $s_{f}$ can be expressed as
\begin{equation}
s_{f} = \frac{L\rho^{\alpha}}{2^{G+1-f}(\rho^{\alpha}-1)}\left( 1+\frac{1}{\rho^{\alpha+\beta(G+1-f)}}\right)\times q\left(\frac{1}{\rho^{\alpha}};\frac{1}{\rho^{\beta}}\right)_{G+1-f}.
\label{s_m1}
\end{equation}
Thus with the knowledge of $s_{f}$, $f=1,2,....,G$ and starting from a rectangular potential barrier of width $l_{G}$ and height $V$, the $G^{th}$ stage UCP-$\rho$ system can be obtained through the following sequence of operation: $V_{1}= (V,2,s_{1})$, then obtain $V_{2}= (V_{1},2,s_{2})$, then $V_{3}= (V_{2},2,s_{3})$$,.....$, $V_{G}= (V_{G-1},2,s_{G})$.  The generated system $V_{G}$ will be our $G^{th}$ stage UCP-$\rho$ system. In the next, we calculate the tunneling probability from UCP-$\rho$ potential system.   
\section{Tunneling probability from UCP-$\rho$ system}
\label{tucp}
In the above section, it has been shown that the UCP-$\rho$ system is a special case of SPP system with periodic repetitions $N_{q}=2$, $q \in \{1,2,3,...,G \}$. If the transfer matrix of one dimensional \textit{initiator} or unit cell potential is known
\beq
 \mathcal{M}(k)= \begin{pmatrix}   \mathcal{M}_{11}(k) & \mathcal{M}_{12}(k) \\ \mathcal{M}_{21}(k) & \mathcal{M}_{22}(k) \end{pmatrix},
\eeq
then the transmission probability of SPP of order $G$ is given by \cite{mh_spp},
\begin{equation}
		T(N_{1},N_{2},....,N_{G})=\frac{1}{1+[|\mathcal{M}_{12}|U_{N_{1}-1}(\Omega_{1})U_{N_{2}-1}(\Omega_{2})U_{N_{3}-1}(\Omega_{3})........U_{N_{G}-1}(\Omega_{G})]^{2}}. 
		\label{t_spp}
	\end{equation}
Where $N_{1}$, $N_{2}$,....,$N_{G}$ is the number of periodicity of the unit cell potential of order $1$, $2$,....,$G$ respectively. $U_{N}(\Omega)$ is the Chebyshev polynomial of second kind \cite{abramowitz1964}. Various $\Omega$s appearing in the above equation are the arguments of Chebyshev polynomial  which represents the Bloch phases of the corresponding fully developed periodic systems. The expression for $\Omega_{q}$, $ q \in \{1,2,3,...,G \}$ is given by \cite{mh_spp, vn_svc},  
\begin{multline}
\Omega_{q}(k) = \vert \mathcal{M}_{22} \vert \cos \left[\theta - k \left \{ \sum_{p=1}^{q-1}(N_{p}-1)s_{p} - s_{q}\right \} \right ] {\prod_{p=1}^{q-1}U_{N_{p}-1}(\Omega_{p})} \\ -\sum_{r=1}^{q-2}\left [\cos \left\{k\left({\sum_{p=r}^{q-1}N_{p}s_{p}} - {\sum_{p=r+1}^{q}s_{p}}\right)\right\}U_{N_{r}-2}(\Omega_{r}){\prod_{p=r+1}^{q-1}U_{N_{p}-1}(\Omega_{p})}\right ]\\-U_{N_{q-1}-2}(\Omega_{q-1})\cos\{k(N_{q-1}s_{q-1}-s_{q})\}.
\label{Omega_nn}
\end{multline}
Where, $\theta$ represents the argument of $\mathcal{M}_{22}(k)$. The expression is also valid for $q$ = $1$ and $2$ provided we drop the terms when the running variable is more than the upper limit for the summation symbols and take the terms as unity when the running variable is more than the upper limit for the product symbols. Also in the above series $N_{0}=1$ and $s_{0}=0$.\\
\indent
As we know for UCP-$\rho$ system, $N_{q}=2$, $ q \in \{1,2,3,...,G \}$ therefore we simplify Eq. (\ref{Omega_nn}) by using $N_{q}= 2$, $U_{0}(z) =1$ and $U_{1}(z) =2z$ to obtain, 
\begin{equation}
\Omega_{q}(k) = 2^{q-1}\vert \mathcal{M}_{22} \vert\cos\big\{\theta-k\gamma_{1}(q)\big\} \prod_{p=1}^{q-1} \Omega_{p} - \sum_{r=1}^{q-1}\left\{2^{q-r-1}\cos \big \{k\gamma_{2}(q,r)\big \} \prod_{p=r+1}^{q-1}\Omega_{p}\right\},
\label{Omega_n1}
\end{equation}
where $\gamma_{1} (q)$ and $\gamma_{2}(q,r)$ is given by,
\begin{equation}
\gamma_{1}(q) = \left\{\sum_{p = 1}^{q-1}s_{p}\right\}-s_{q},
\label{eta1}
\end{equation}
\begin{equation}
\gamma_{2}(q,r)  =\left\{\sum_{p = r}^{q}s_{p}\right\}- ( 2s_{q}-s_{r}).
\end{equation}
It can be easily shown that
\begin{equation}
\gamma_{2}(q,r)  =\gamma_{1}(q) -\gamma_{1}(r).
\label{gamma2}
\end{equation}
In Eq. (\ref{Omega_n1}), when the running variable is greater than the upper limit for the summation symbol then the term will be dropped out and when the running variable is greater than the upper limit of the product symbol then the terms will be unity.\\
\indent
Next, let us find the simplified expression for $\gamma_{1}(q)$ and $\gamma_{2}(q, r)$. Fig. \ref{general_figure} implies that
\begin{equation}
   s_{1} = l_{G} + d_{G}, \nonumber
\end{equation}
\begin{equation}
    s_{2} = s_{1} + l_{G} + d_{G-1}, \nonumber
\end{equation}
\begin{equation}
    s_{3} = s_{1} + s_{2} + l_{G} + d_{G-2}, \nonumber
\end{equation}
\begin{equation}
    s_{4} = s_{1} + s_{2} + s_{3} + l_{G} + d_{G-3}. \nonumber
\end{equation}
In general, the periodic distance for each stage $G$ can be expressed through
\begin{equation}
s_{f} = \left(\sum_{p = 1}^{f-1}s_{p}\right)+l_{G} + d_{G-f+1}.
\end{equation}
Using above equation in Eq. (\ref{eta1}), $\gamma_{1}(q)$ is expressed through,
\begin{equation}
    \gamma_{1}(q) = -(l_{G} + d_{G-q+1}),
\label{gamma1}
\end{equation}
which implies that $\gamma_{1}(q)$ is always negative. Now using Eq. (\ref{gamma1}) in Eq. (\ref{gamma2}), we get
\begin{equation}
    \gamma_{2}(q, r) = d_{G-r+1} - d_{G-q+1}.
    \label{gamma22}
\end{equation}
It can be seen in Fig. \ref{general_figure} that for $a>b$, $d_{a}<d_{b}$. Also for $r<q$, $G-r+1>G-q+1$. Hence, it is clear that $d_{G-r+1}<d_{G-q+1}$. Therefore, for $r<q$, $\gamma_{2}(q, r) < 0$. Eq. (\ref{gamma1}) and (\ref{gamma22}) gives the general expression for $\gamma_{1}(q)$ and $\gamma_{2}(q, r)$ respectively.\\
\indent
The above discussion completes the calculation for $\Omega_{q}$. With the knowledge of  $\Omega_{1}$, $\Omega_{2}$, $\Omega_{3}$,....,$\Omega_{G}$ one can find the transmission probability by using Eq. (\ref{t_spp}) with $N_{q}=2$. The transfer matrix of `unit cell' rectangular barrier of width $l_{G}$ and height $V$ is,\\
\begin{subequations}
	\begin{equation}
			\mathcal{M}_{11}=(\cos{\kappa l_{G}}-i\epsilon_{+} \sin{\kappa l_{G}})e^{ikl_{G}},
	\end{equation}
    \begin{equation}
			\mathcal{M}_{12}=i\epsilon_{-} \sin{\kappa l_{G}},
	\end{equation}		
	\begin{equation}
	  	\mathcal{M}_{21}=-i\epsilon_{-} \sin{\kappa l_{G}},
	\end{equation}
		\begin{equation}
			\mathcal{M}_{22}=(\cos{\kappa l_{G}}+i\epsilon_{+} \sin{\kappa l_{G}})e^{-ikl_{G}}.
		\end{equation}
\end{subequations}
	\begin{equation}
		\epsilon_{\pm}=\frac{1}{2} \Big( \tau\pm \frac{1}{\tau}\Big), 
		\label{epsilon_plus_minus}
	\end{equation}
	\begin{equation}
		\tau=\frac{k}{\kappa} \ \ , \ \ k=\frac{\sqrt{2mE}}{\hbar}  \nonumber 
	\end{equation} 
	and
	\begin{equation}
		\kappa=\frac{\sqrt{2m(E-V)}}{\hbar}. \nonumber
	\end{equation}
With $N_{q}=2$ and using the knowledge of the transfer matrix of unit cell rectangular barrier, simplify Eq. (\ref{t_spp}) to obtain the final expression of transmission coefficient from UCP-$\rho$ system as,
\begin{equation}
T_{G}(k)=\frac{1}{1+4^{G}\epsilon_{-}^{2}\sin^{2}{(\kappa l_{G})} \prod_{q=1} ^{G} \Omega_{q}^{2}}.
\label{T_general_rho}
\end{equation}
In the next subsections, we obtained the expression for $\gamma_{1}(q)$ and $\gamma_{2}(q,r)$ for general Cantor potential SVC-$\rho$ potential system which are the special cases of UCP-$\rho$ system.
\subsection{Case 1: General Cantor potential system}
\label{gc}
General Cantor fractal system is an special case of UCP-$\rho$ system. In this case, $\alpha = 1$ and $\beta = 0$. Hence, from Eq. (\ref{general_lG}), length $l_{G}$ of the potential segment at any arbitrary stage G for general Cantor fractal system is expressed through
\begin{equation}
l_{G} = L\left(\frac{\rho - 1}{2\rho}\right)^{G}.
\label{lg1}
\end{equation}
For the general Cantor potential system, we remove a fraction $1/\rho$ from the middle of each potential segment of stage $G-1$, therefore $d_{G} = l_{G-1}/\rho$. Substituting here for $l_{G}$, we get
\begin{equation}
    d_{G} = \frac{L}{\rho}\left(\frac{\rho-1}{2\rho}\right)^{G-1}.
    \label{dg1}
\end{equation}
Using Eq. (\ref{lg1}) and (\ref{dg1}) in Eq. (\ref{gamma1}), we get $\gamma_{1}(q)$ for general Cantor system as 
\begin{equation}
    \gamma_{1}(q) = -L\left(\frac{\rho-1}{2\rho}\right)^{G}\left\{1-\frac{1}{\rho}\left(\frac{2\rho}{\rho-1}\right)^{q}\right\}.
\end{equation}
Using above equation in Eq. (\ref{gamma22}) and after some simplification, we get $\gamma_{2}(q, r)$ for general Cantor system as
\begin{equation}
    \gamma_{2}(q, r) = \frac{L}{\rho}\left\{\left(\frac{\rho-1}{2\rho}\right)^{G-r}-\left(\frac{\rho-1}{2\rho}\right)^{G-q}\right\}.
\end{equation}
From Eq. (\ref{dg1}), the above equation is also expressed as
\begin{equation}
   \gamma_{2}(q, r) = d_{G-r+1}-d_{G-q+1}.
\end{equation}
For $\rho = 3$, whole system behaves as \textit{standard} Cantor fractal potential system and in this case $l_{G} = d_{G} = L/3^{G}$, and $\gamma_{1}(q)$ and $\gamma_{2}(q, r)$ are expressed through
\begin{equation}
    \gamma_{1}(q) = -\frac{L}{3^{G}}\left(1-\frac{1}{3^{1-q}}\right).
\end{equation}
and
\begin{equation}
    \gamma_{2}(q, r) = \frac{L}{3^{G-r+1}}\left(1-\frac{1}{3^{r-q}}\right).
\end{equation}
respectively.
With the knowledge of the above mentioned parameters, we can express the Bloch phase for the general and \textit{standard} Cantor fractal system by using Eq. (\ref{Omega_nn}) and then tunneling probability by using Eq. (\ref{T_general_rho}) and this is also discussed in \cite{mh_spp}.
\subsection{Case 2: General Smith-Volterra-Cantor system}
\label{gsvc}
General Smith-Volterra-Cantor (GSVC or SVC-$\rho$) is also an special case of UCP-$\rho$ system. In this case, $\alpha = 0$ and $\beta = 1$. Substituting this value of $\alpha$ and $\beta$ in Eq. (\ref{general_lG}), $l_{G}$ for general SVC is expressed through
\begin{equation}
l_{G} = \frac{L}{2^{G}}\prod_{j=1}^{G}\left(1-\frac{1}{\rho^{j}}\right).
\label{lg2}
\end{equation}
For SVC-$\rho$ potential system, we remove a fraction $1/\rho^{G}$ from the middle of each potential segment of stage $G-1$, therefore $d_{G} = l_{G-1}/\rho^{G}$. Substituting here for $l_{G}$, we get
\begin{equation}
    d_{G} = \frac{L}{\rho^{G}2^{G-1}}\prod_{j=1}^{G-1}\left(1-\frac{1}{\rho^{j}}\right).
    \label{dg2}
\end{equation}
Using Eq. (\ref{lg2}) and (\ref{dg2}) in Eq. (\ref{gamma1}), we get $\gamma_{1}(q)$ for SVC-$\rho$ system as
\begin{align}
    \gamma_{1}(q) = -\left\{\frac{L}{2^{G}}\prod_{j=1}^{G}\left(1-\frac{1}{\rho^{j}}\right)+\frac{L}{2^{G-q}\rho^{G-q+1}}\prod_{j=1}^{G-q}\left(1-\frac{1}{\rho^{j}}\right)\right\}.
\end{align}
Using above equation in Eq. (\ref{gamma2}) and after some simplification, we get $\gamma_{2}(q, r)$ for SVC-$\rho$ system as
\begin{equation}
    \gamma_{2}(q,r) = \frac{L}{2^{G-r}\rho^{G-r+1}}\prod_{j=1}^{G-r}\left(1-\frac{1}{\rho^{j}}\right)-\frac{L}{2^{G-q}\rho^{G-q+1}}\prod_{j=1}^{G-q}\left(1-\frac{1}{\rho^{j}}\right).
\end{equation}
Using Eq. (\ref{dg2}) in above equation, $\gamma_{2}(q,r)$ is expressd through
\begin{equation}
    \gamma_{2}(q,r) = d_{G-r+1}-d_{G-q+1}.
\end{equation}
For $\rho = 4$, whole system behaves as \textit{standard} SVC (SVC-4) potential system and in this case $l_{G}$, $d_{G}$, $\gamma_{1}(q)$ and $\gamma_{2}(q, r)$ are expressed through
\begin{equation}
l_{G} = \frac{L}{2^{G}}\prod_{j=1}^{G}\left(1-\frac{1}{4^{j}}\right),
\end{equation}
\begin{equation}
    d_{G} = \frac{2L}{8^{G}}\prod_{j=1}^{G-1}\left(1-\frac{1}{4^{j}}\right),
\end{equation}
\begin{equation}
     \gamma_{1}(q) = -\left\{\frac{L}{2^{G}}\prod_{j=1}^{G}\left(1-\frac{1}{4^{j}}\right)+\frac{L}{4\times8^{G-q}}\prod_{j=1}^{G-q}\left(1-\frac{1}{4^{j}}\right)\right\}
\end{equation}
and
\begin{align}
   \gamma_{2}(q,r) = \frac{L}{4\times8^{G-r}}\prod_{j=1}^{G-r}\left(1-\frac{1}{4^{j}}\right)-\frac{L}{4\times8^{G-q}}\prod_{j=1}^{G-q}\left(1-\frac{1}{4^{j}}\right) = d_{G-r+1}-d_{G-q+1}
\end{align}
respectively. Again, with the knowledge of the above mentioned parameters we can express the Bloch phase for the SVC-$\rho$ and SVC-4 potential system by using Eq. (\ref{Omega_nn}) and then tunneling probability by using Eq. (\ref{T_general_rho}) and this is also reported in \cite{vn_svc}.
\section{Transmission features}
\label{transfeatures}
\begin{figure}[h! tbp]
\begin{center}
     \includegraphics[scale=0.35]{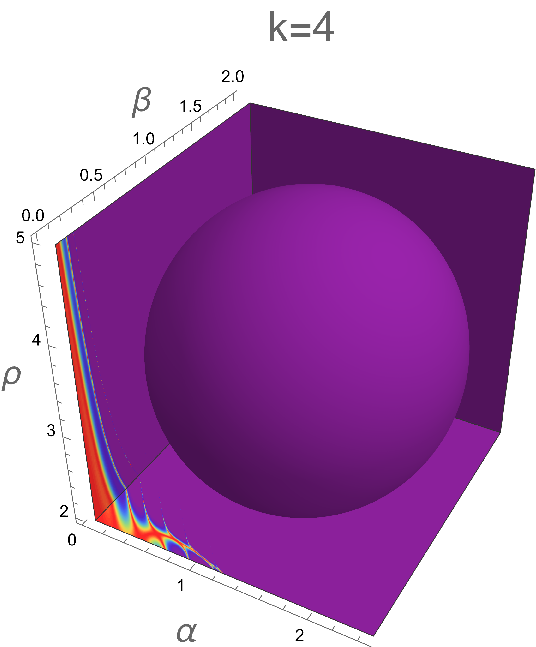}(a)
    \includegraphics[scale=0.35]{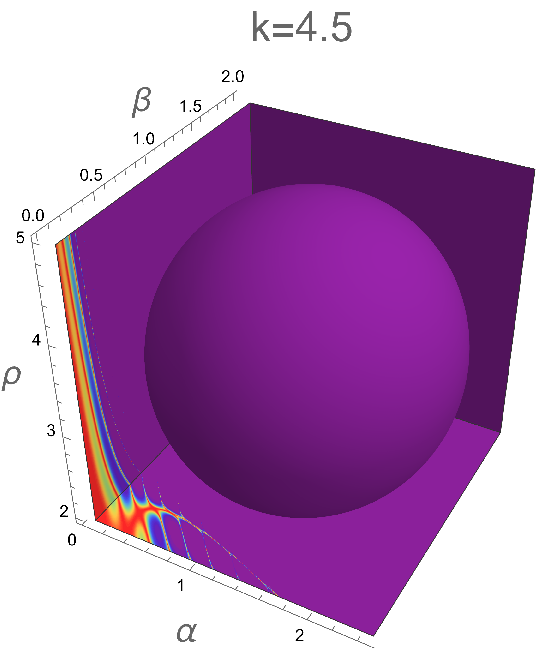}(b)
     \includegraphics[scale=0.35]{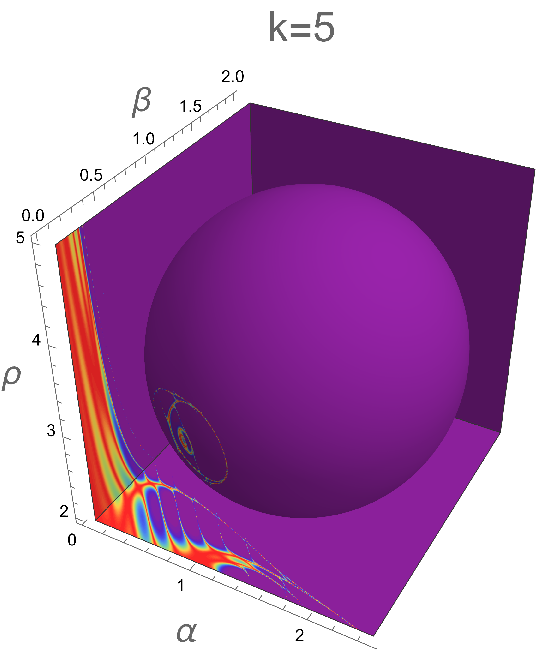}(c)\includegraphics[scale=0.43]{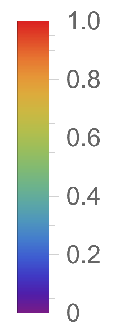}\\
     \includegraphics[scale=0.35]{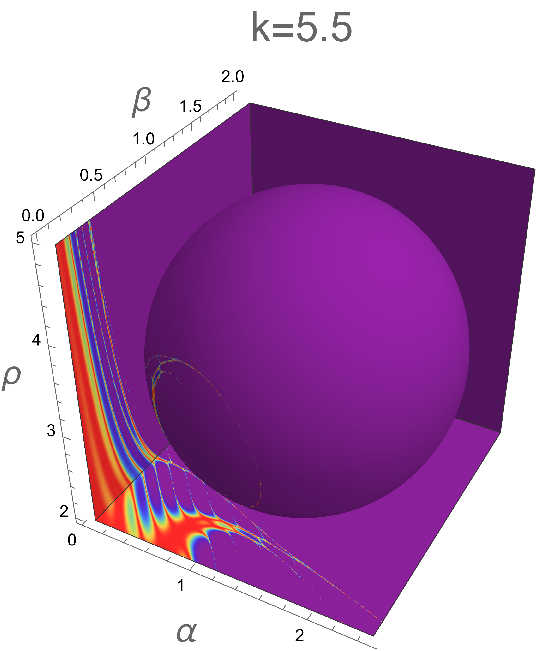}(d)
     \includegraphics[scale=0.35]{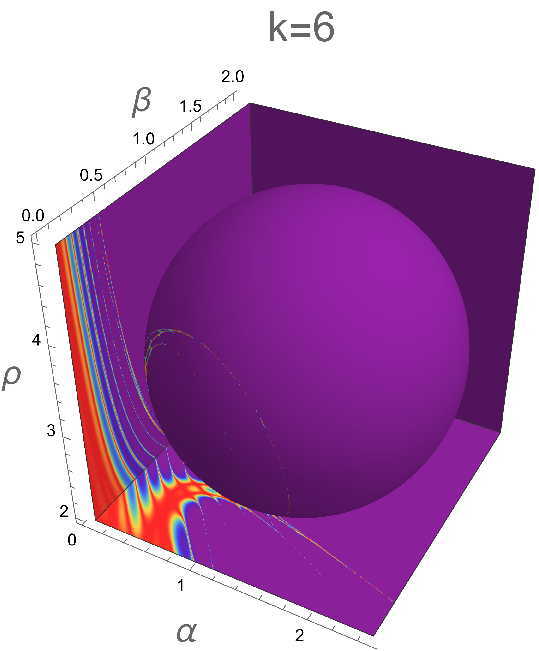}(e)
     \includegraphics[scale=0.35]{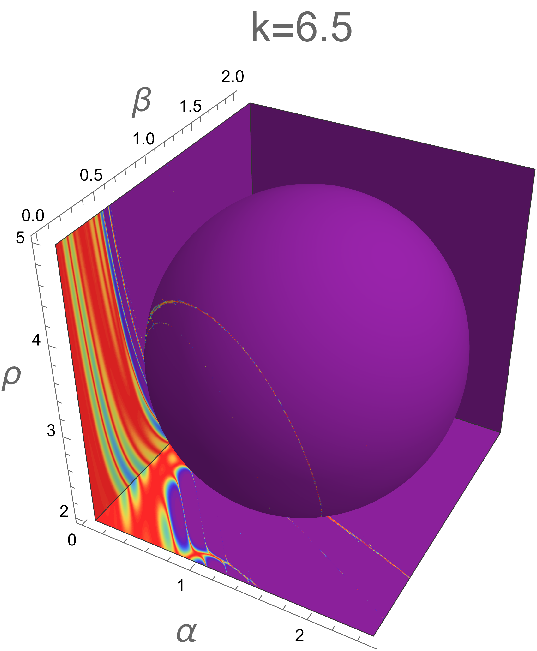} (f) \includegraphics[scale=0.43]{barlegend.png}\\
     \includegraphics[scale=0.35]{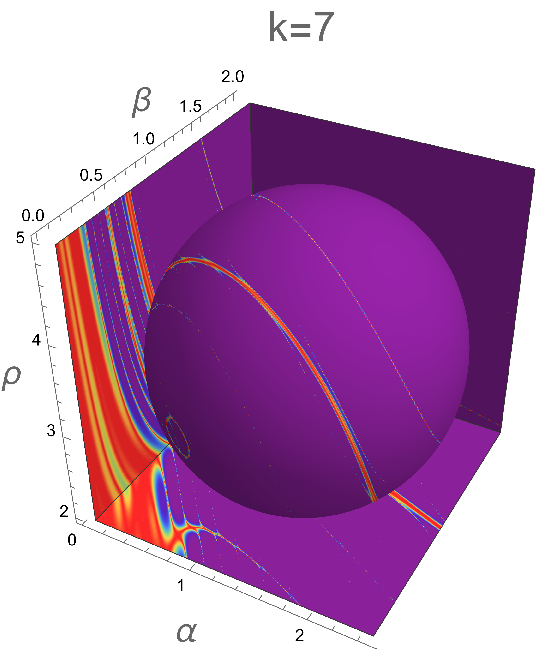}(g)
     \includegraphics[scale=0.35]{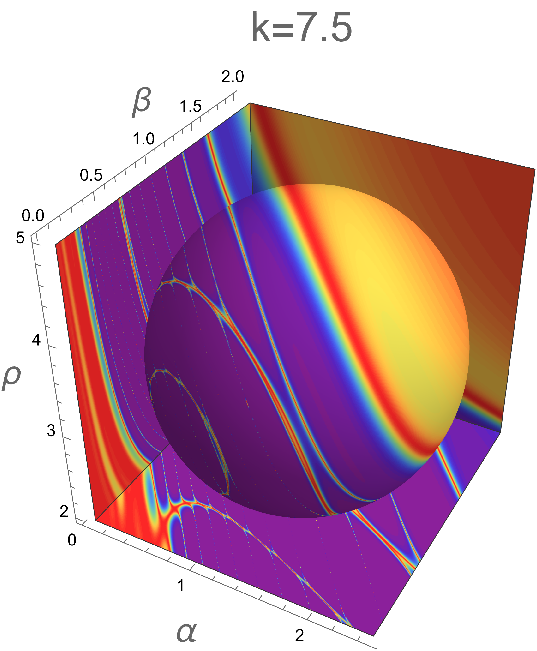}(h)
     \includegraphics[scale=0.35]{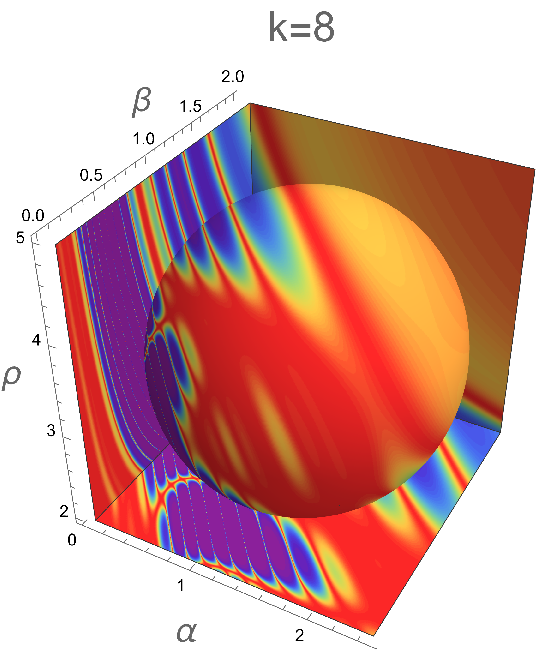}(i)
     \includegraphics[scale=0.425]{barlegend.png}
    \caption{\it Figure shows the graphical representation of tunneling probability $T$ for UCP-$\rho$ potential of stage $G=3$, $L=10$ and $V=25$ in the parameter space of $\alpha$, $\beta$ and $\rho$ for different values of wave vector $k$. The figures shows the density plots of $T$ in the backplanes as well as in the centre sphere of cuboid spanned by finite range of $\alpha$, $\beta$ and $\rho$. Very sharp resonances are noticed from this potential.}
   \label{UF1}
\end{center}    
\end{figure}

\begin{figure}[h! tbp]
\begin{center}
    \includegraphics[scale=0.35]{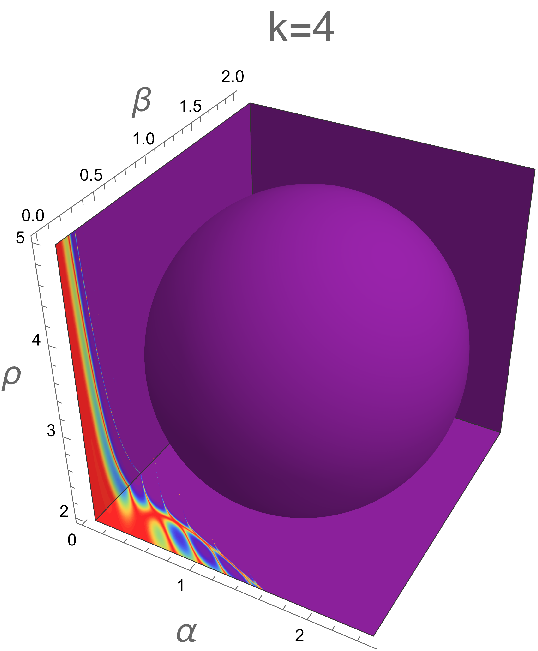}(a)
     \includegraphics[scale=0.35]{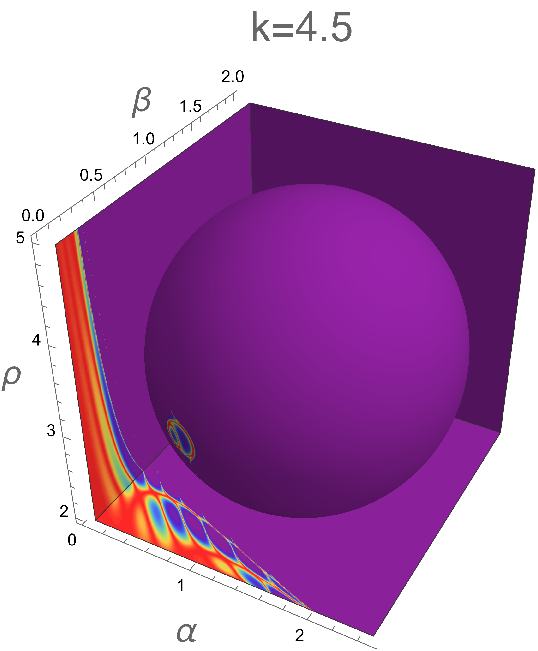}(b)
     \includegraphics[scale=0.35]{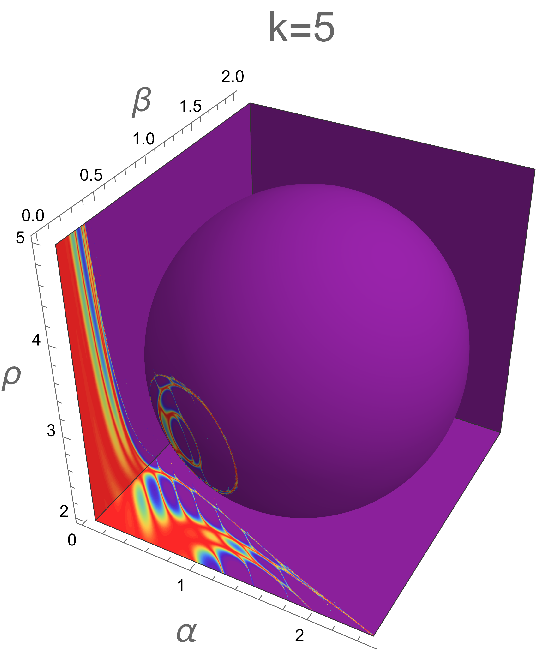}(c)\includegraphics[scale=0.43]{barlegend.png}\\
     \includegraphics[scale=0.35]{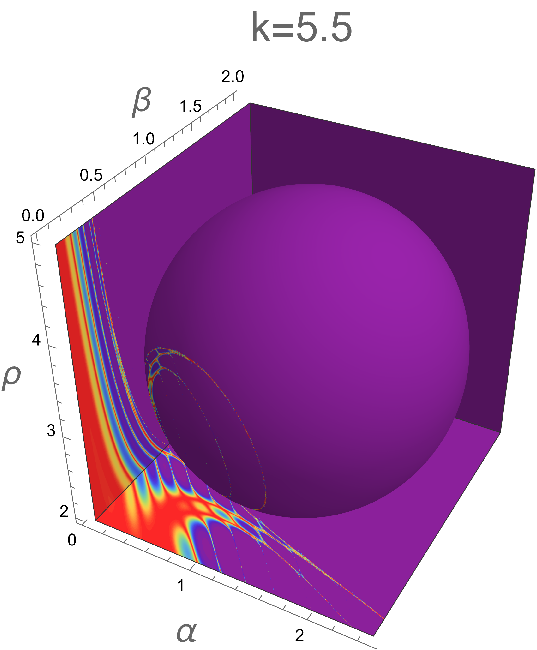}(d)
     \includegraphics[scale=0.35]{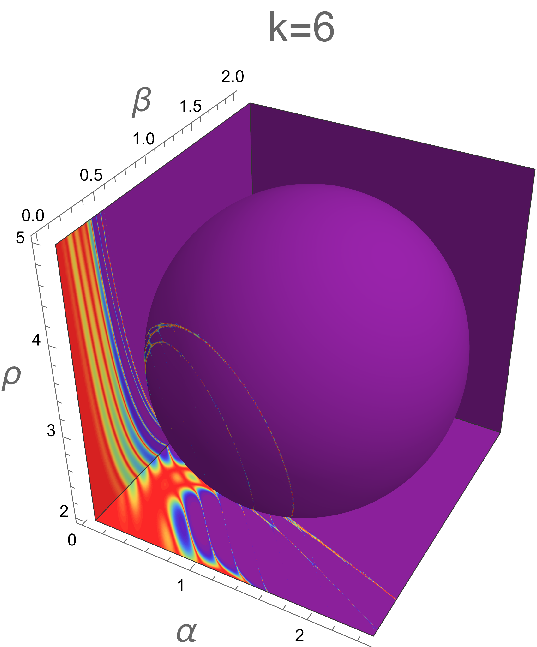}(e)
     \includegraphics[scale=0.35]{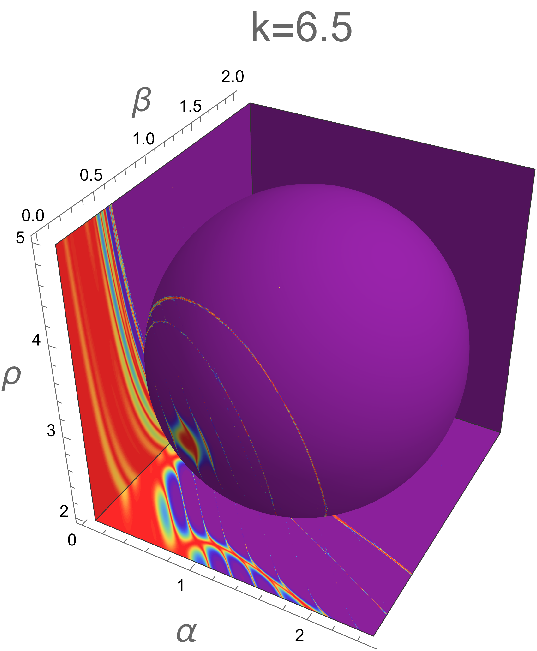}(f)\includegraphics[scale=0.43]{barlegend.png}\\
     \includegraphics[scale=0.35]{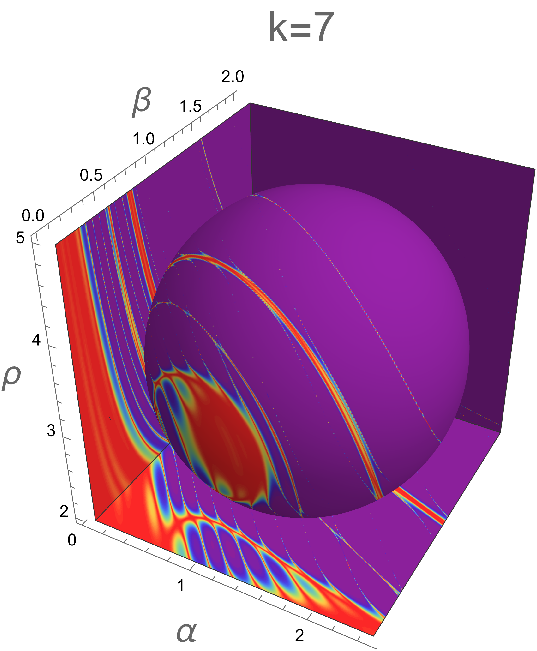}(g)
     \includegraphics[scale=0.35]{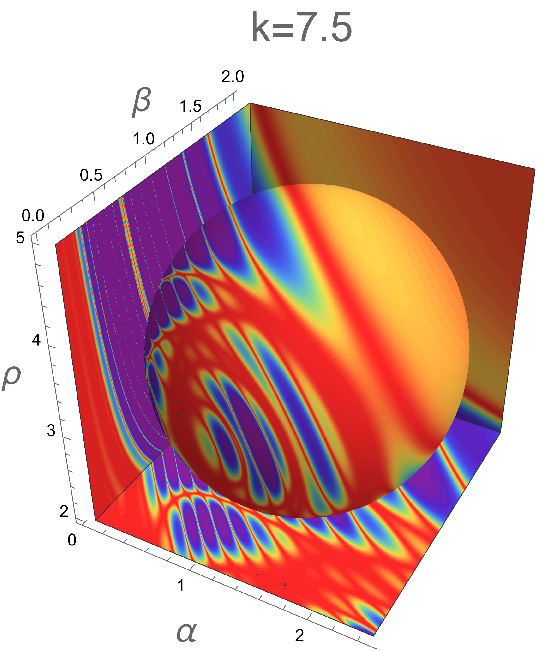}(h)
     \includegraphics[scale=0.35]{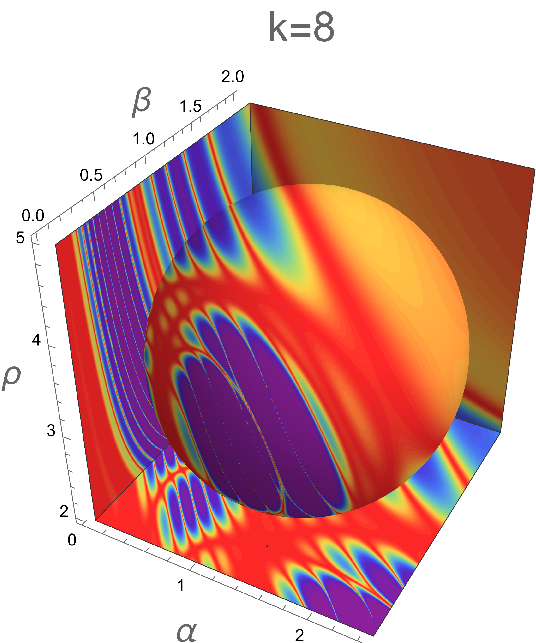}(i)
     \includegraphics[scale=0.43]{barlegend.png}
    \caption{\it Figure shows the graphical representation of tunneling probability $T$ for UCP-$\rho$ potential of stage $G=4$, $L=10$ and $V=25$ in the parameter space of $\alpha$, $\beta$ and $\rho$ for different values of wave vector $k$. The figures shows the density plots of $T$ in the backplanes as well as in the centre sphere of cuboid spanned by finite range of $\alpha$, $\beta$ and $\rho$. Very sharp resonances are noticed from this potential.}
    \label{UF2}
\end{center}    
\end{figure}

In this section, we present several interesting features of tunneling from UCP-$\rho$ system. Fig. \ref{UF1} and \ref{UF2} shows the graphical representation of tunneling probability $T$ for stage $G=3$, $4$ respectively in the parameter space of $\alpha$, $\beta$ and $\rho$ for different values of wave vector $k$. For both the figures $L=10$, $V=25$. The figures shows the density plots of $T$ in the backplanes as well as in the centre sphere of the cuboid spanned by finite range of $\alpha$, $\beta$ and $\rho$ as shown in the figure. Both the figures show very sharp transmission resonances.    

\subsection{Scaling relation of reflection coefficient}
\label{scaling}
\begin{figure}[H]
\begin{center}
     \includegraphics[scale=0.452]{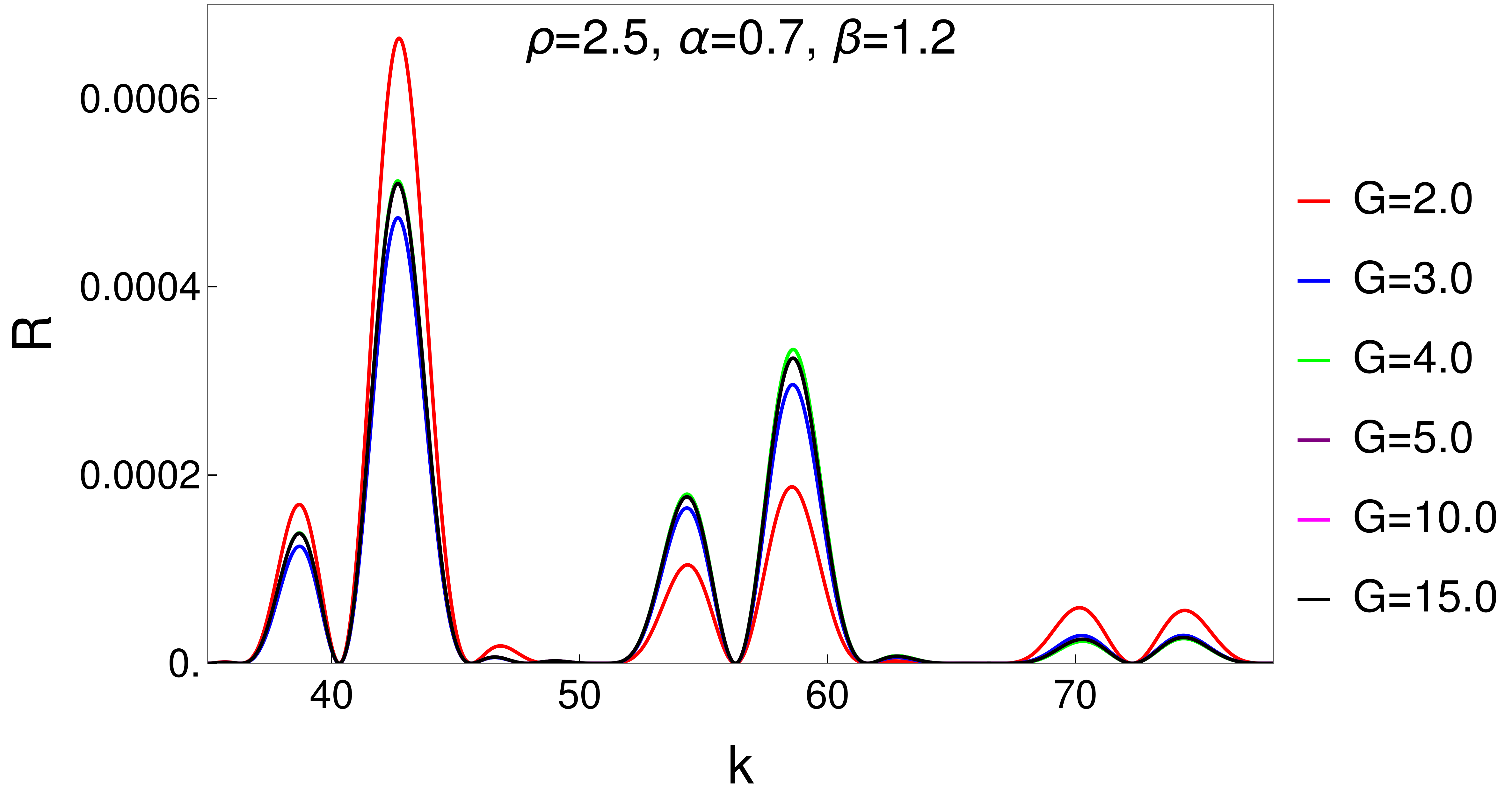}(a)\\
     \includegraphics[scale=0.452]{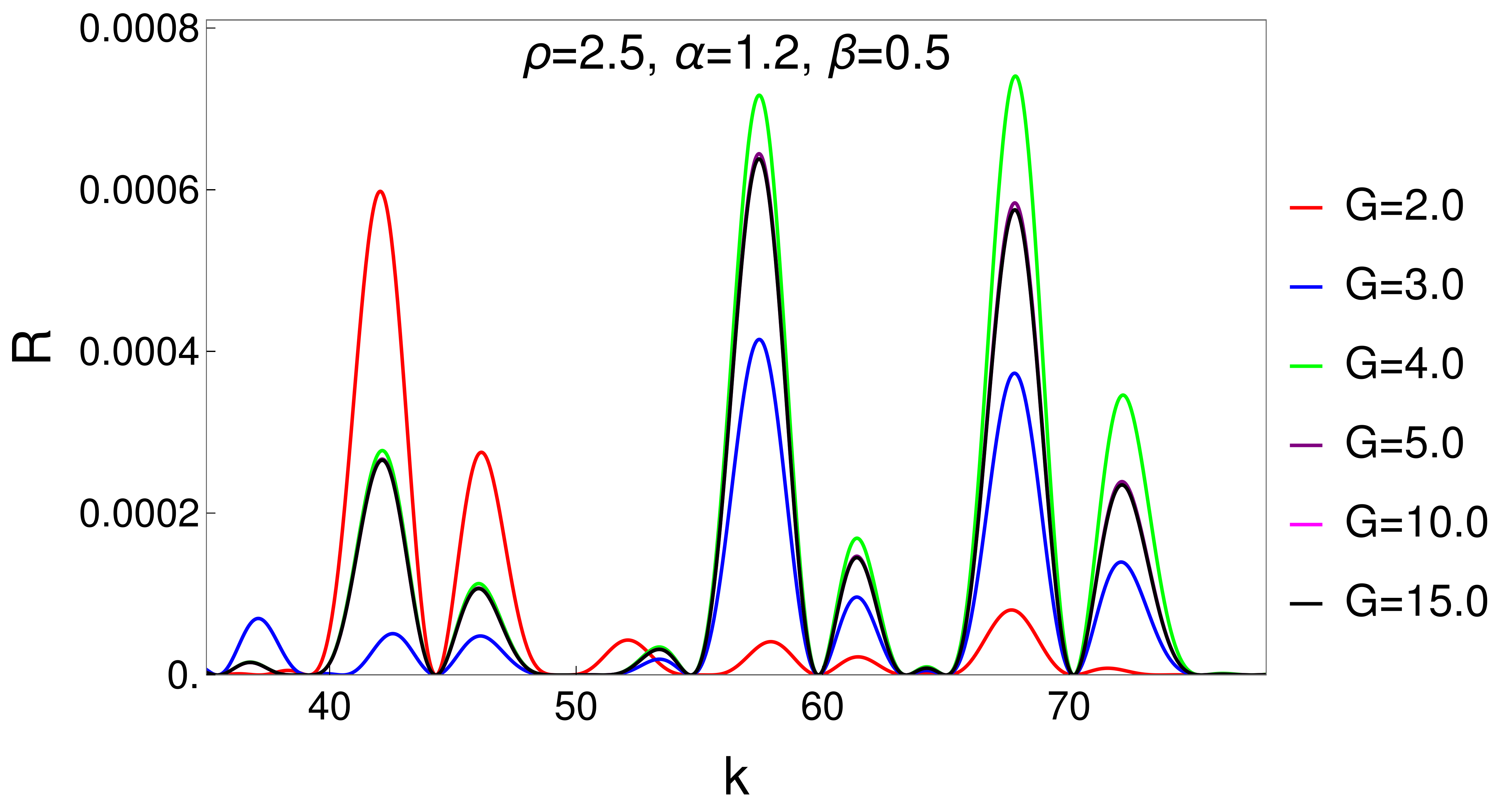}(b)
    \caption{\it{Plots showing the variation of $R_{G}(k)$ for UCP-$\rho$ system for different stages $G$. The potential height $V_{G}$ is determined from Eq. (\ref{vg_values}). Here potential parameters are $L=1$, $V_{0}=10$, and other parameters are mentioned in plots. From the above plots, we observe that reflection peaks are well separated for lower stage values i.e. for $G$= 2, 3 and 4. However, as stage $G$ attains higher values (i.e. $G$=5, 10 and 15) the difference between reflection profiles are invisible. This shows the convergence of the product term of Eq. (\ref{vg_zetag}) } }
    \label{UF_saturation}
\end{center}    
\end{figure}

\begin{figure}[H]
\begin{center}
    \includegraphics[scale=0.35]{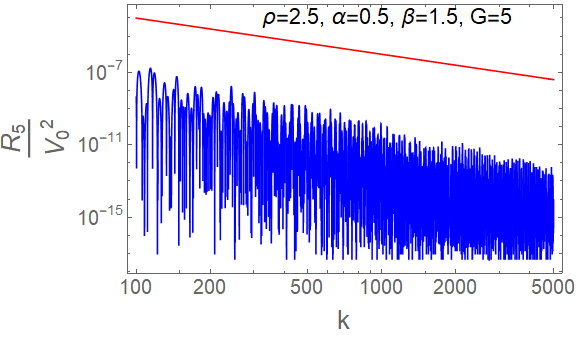} (a)
    \includegraphics[scale=0.35]{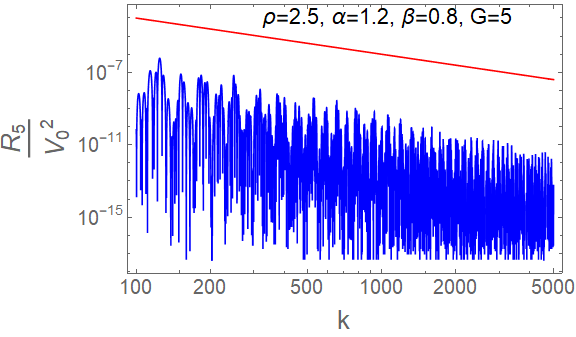} (b)\\
    \includegraphics[scale=0.35]{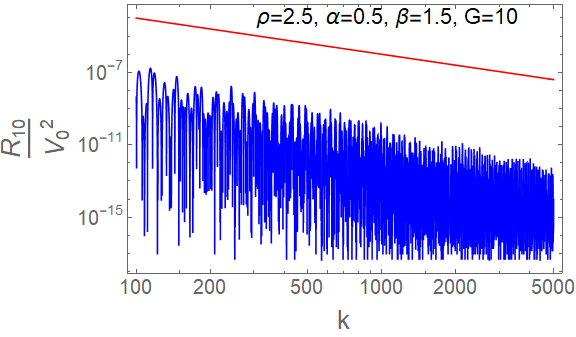} (c)
    \includegraphics[scale=0.35]{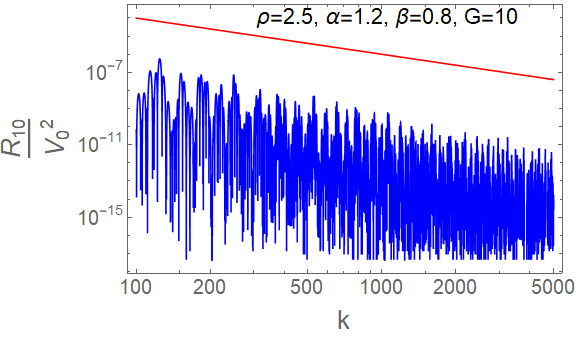} (d)\\
    \includegraphics[scale=0.35]{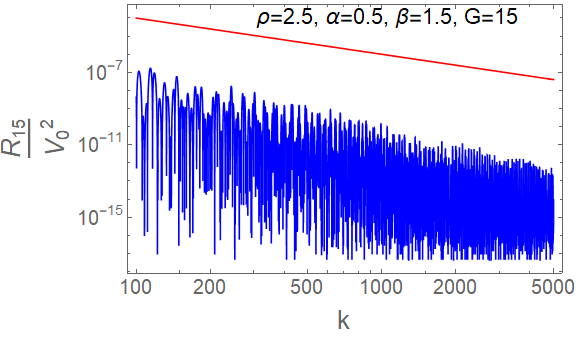} (c)
    \includegraphics[scale=0.35]{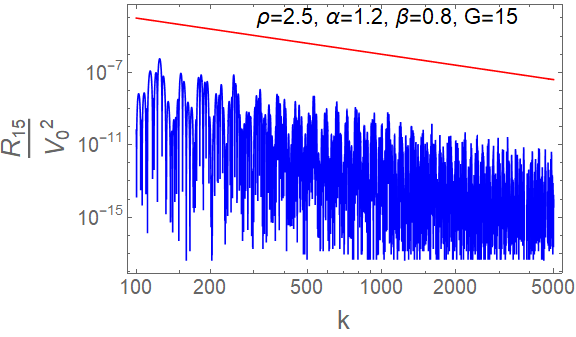} (d)
     \caption{\it{log-log plots showing the scaling behavior of reflection coefficient $\frac{R_{G}}{V_{0}^{2}}$ for large k for two set of values of $\alpha$ and $\beta$ as mentioned in plots. Here potential parameters are $L$=1 and $V_{0}$=10. The red dashed curve represents $\frac{1}{k^{2}}$. It is observed that at large k, $R_{G}$ falls of according to $\frac{1}{k^{2}}$.}}
    \label{UF_Scaling}
\end{center}
\end{figure}

For large $k$, the reflection coefficient $R(k)$ is very small. With this limit, $R(k)$ can be approximated by (\ref{T_general_rho}) as
\begin{equation}
    R(k) \sim 4^{G} \vert  \mathcal{M}_{12} \vert ^{2} \prod_{i=1}^{G} \Omega_{i}^{2}.
    \label{r_small_value}
\end{equation}
Again for larger $k$ we have $\frac{V}{k^2} < < 1$ and upon Taylor expanding, it can be shown in the first order that
\begin{equation}
    \vert \mathcal{M}_{12} \vert ^{2} \sim \left ( \frac{V l_{G}}{2}\right )^{2} \frac{1}{k ^{2}}. 
    \label{m12_small_value}
\end{equation}
Therefore, the expression for $R(k)$ becomes
\begin{equation}
    R(k) \sim 4^{G} \left ( \frac{V l_{G}}{2}\right )^{2} \frac{1}{k ^{2}} \prod_{i=1}^{G} \Omega_{i}^{2}.
    \label{r_small_value}
\end{equation}
If $V_{G}$ is the height of the potential at each stage $G$, then it can be shown that the following value of $V_{G}$ keeps the total area of potential barrier (sum of the area of all potential segment at stage $G$) as constant,
\begin{equation}
    V_{G} =\frac{L}{2^{G} l_{G}} V_{0},
    \label{vg_values}
\end{equation}
where $V_{0}$ is the height of the potential barrier at $G=0$ and $l_{G}$ is the width of unit cell rectangular barrier defined already (Eq. (\ref{general_lg})). If $R_{G}(k)$ is the reflection coefficient at each stage $G$ with a potential height of each segment as $V_{G}$ then, it can be shown that (valid for large $k$)
\begin{equation}
    \frac{R_{G}(k)}{L^{2} V_{0}^{2}} \sim  \left ( \frac{1}{2}  \right )^{2} \frac{1}{k ^{2}} \prod_{i=1}^{G} \Omega_{i}^{2}.
    \label{vg_zetag}
\end{equation}
In Fig. \ref{UF_saturation}, we show the behavior of $R_{G}(k)$ with different stage $G$ of UCP-$\rho$ potential. The difference between the $R_{G}(k)$ profile becomes invisible for higher stage $G$ which suggests fast convergence of product term of Eq. (\ref{vg_zetag}). For standard Cantor potential, this result has been shown already in an earlier work \cite{cantor_f7}. Again, due to the convergence nature of the product term (provided it is evaluated at $V_{G}$) with increasing $G$, it is evident from Eq. (\ref{vg_zetag}) that $R_{G}(k)$ would scale as $\frac{1}{k ^{2}}$ for large $G$ and $k$ values. This is already noted for standard Cantor potential \cite{cantor_f7}. The scaling behavior of $R_{G}$ with $k$ is shown graphically for UCP-$\rho$ potential system in Fig. \ref{UF_Scaling}.
\subsection{Saturation of the transmission coefficient with stage $G$}
\label{saturation}
From the definition of the UCP-$\rho$ potential system, a portion of $\frac{1}{\rho^{\alpha+\beta G}}$ is taken out from the middle of each potential segment at every stage $G$. This would imply that for fixed $\alpha $ and fixed $\beta >0$, a progressively thinner portion will be removed from previous segments as $G$ increases. This also indicates that with increasing stage $G$, the entire potential system would apparently show a saturation with increasing $G$. It is to be noted that the saturation will occur in the total length of potential segments but potentials at different stages will be mathematically different. A consequence of progressively thinner and thinner removal of the potential segments with increasing $G$ will be the saturation of transmission profile $T_{G}(k)$ with $k$ at higher stage $G$. 

\begin{figure}[H]
\begin{center}
    \includegraphics[scale=0.25]{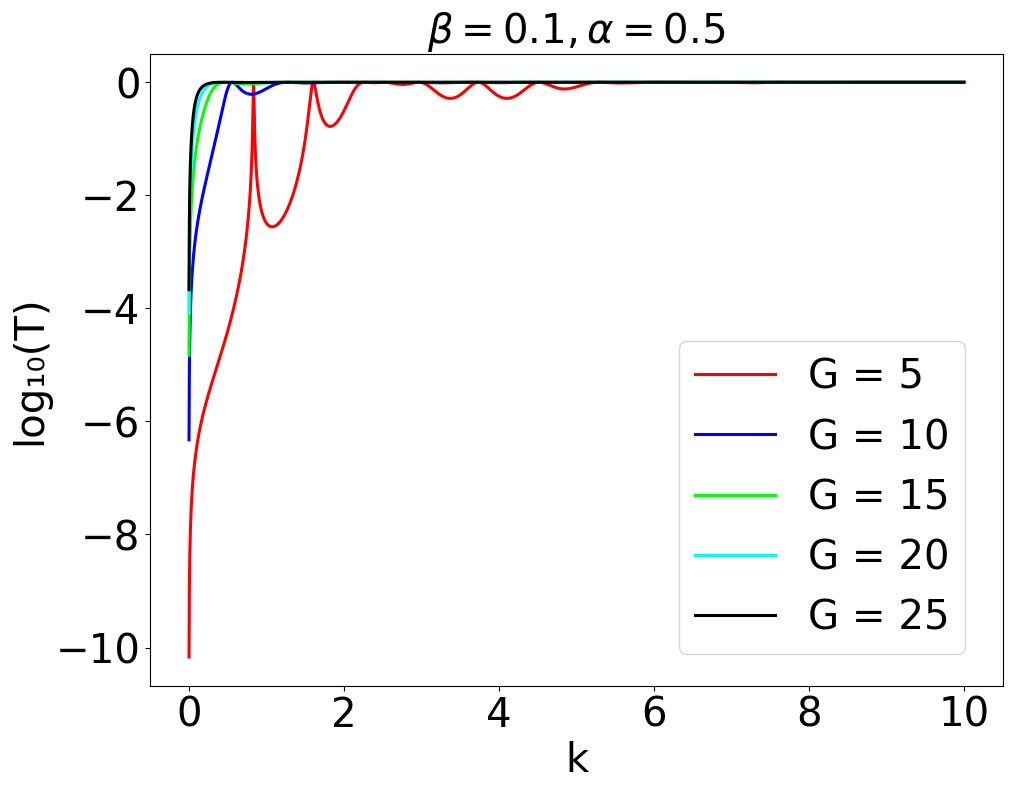} (a)
    \includegraphics[scale=0.25]{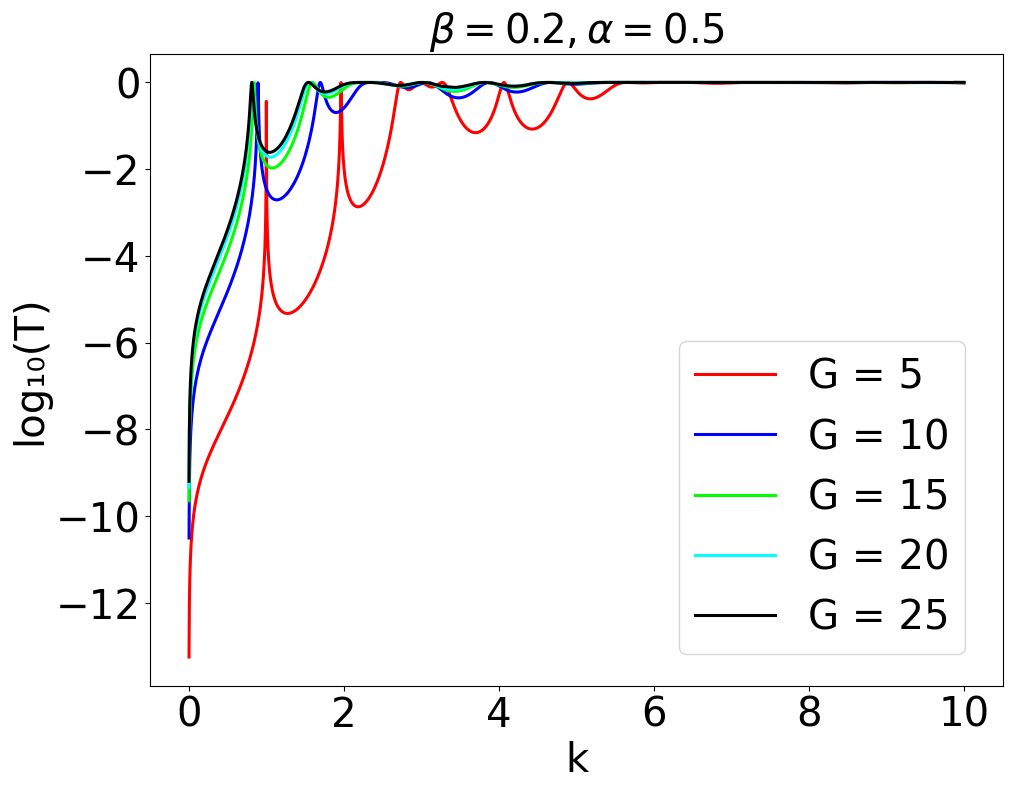} (b)\\
    \includegraphics[scale=0.25]{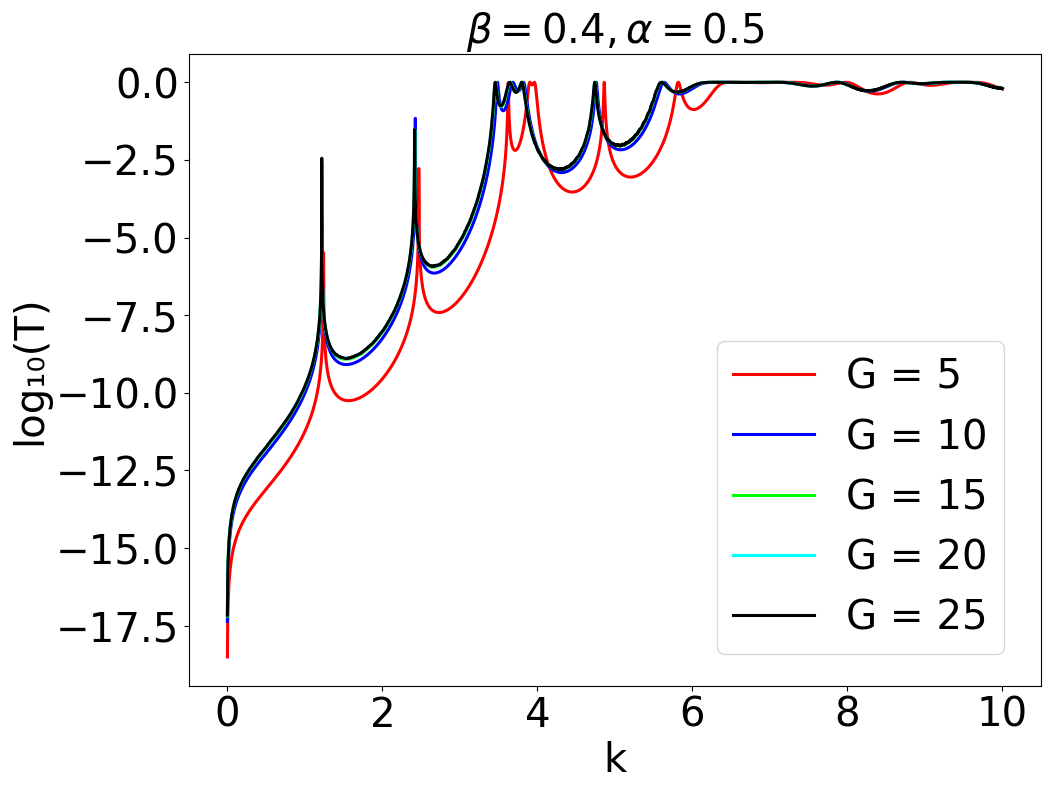} (c)
    \includegraphics[scale=0.25]{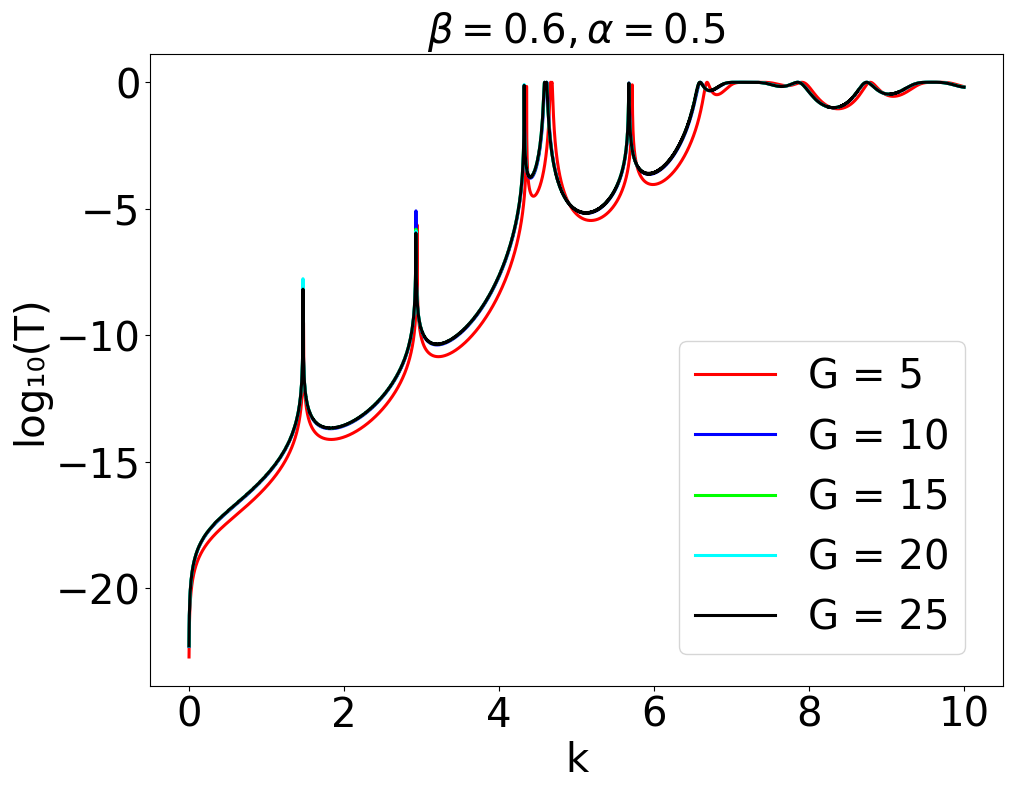} (d)\\
    \includegraphics[scale=0.25]{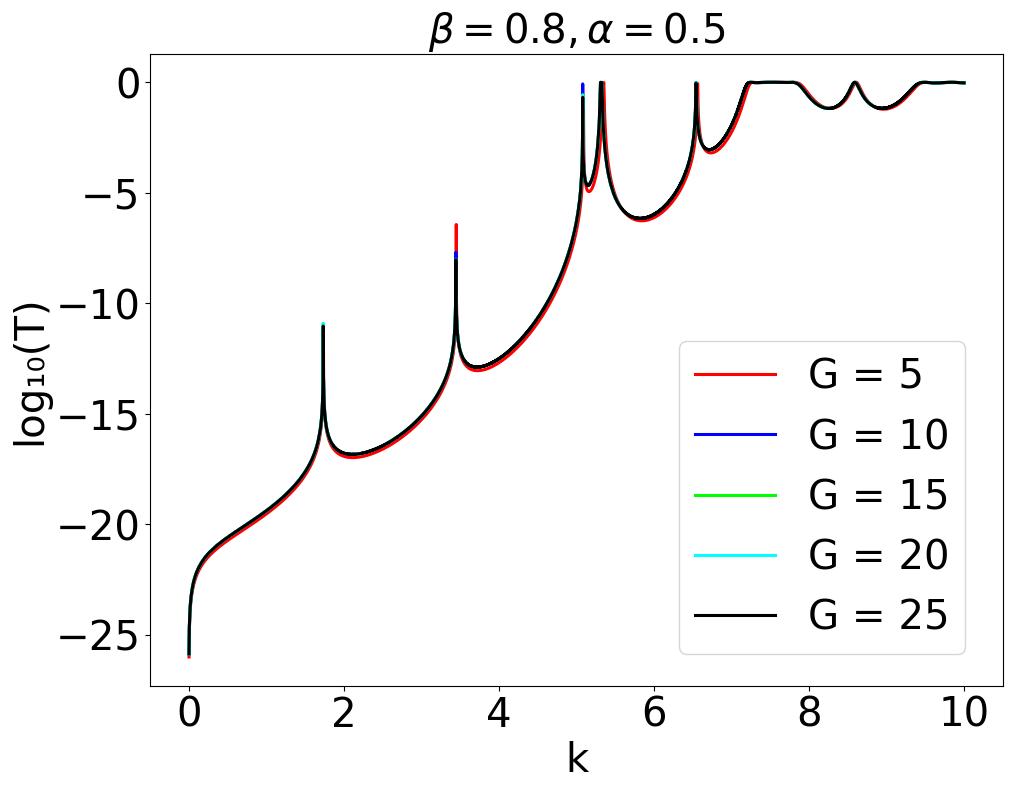} (e)
    \includegraphics[scale=0.25]{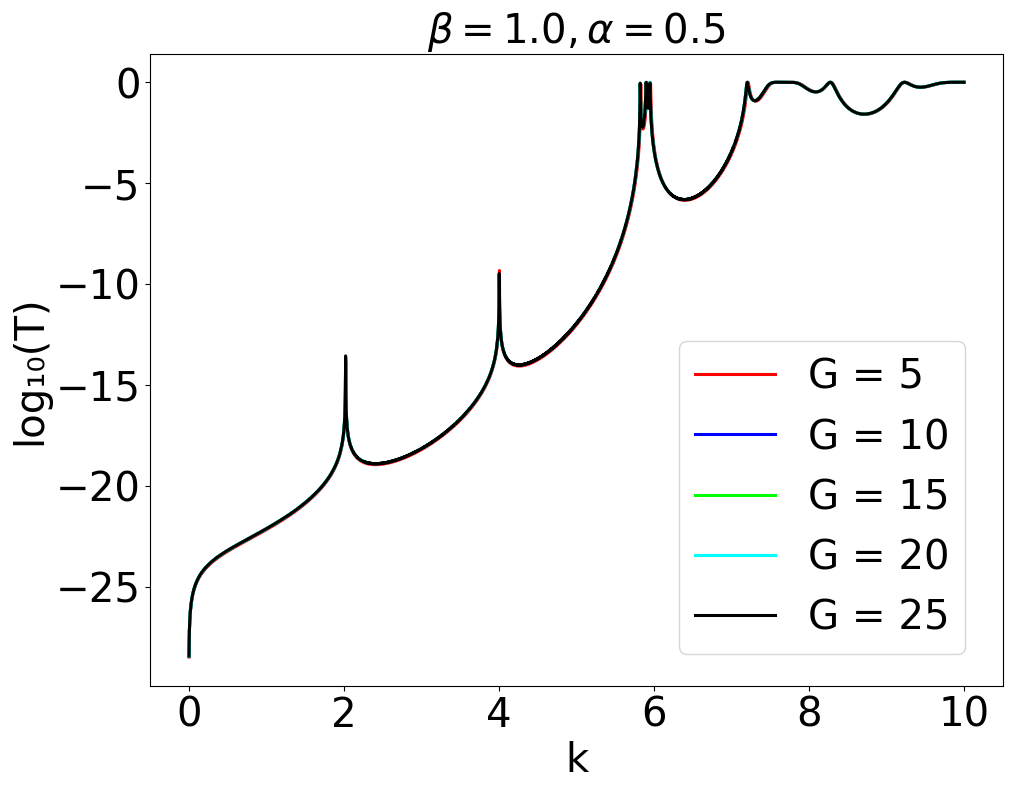} (f)
     \caption{\it{Plots showing the saturation of transmission profile with increasing stage $G$. Here the potential parameters are $L=5$, $V=25$, $\rho= 2.5$ and $\alpha=0.5$. Value of $\beta$ are shown in the figures above. It is seen that for larger $\beta$, saturation occurs for comparatively smaller $G$ values.}}
    \label{saturationplot}
\end{center}
\end{figure}

\begin{figure}[h! tbp]
\begin{center}
    \includegraphics[scale=0.2]{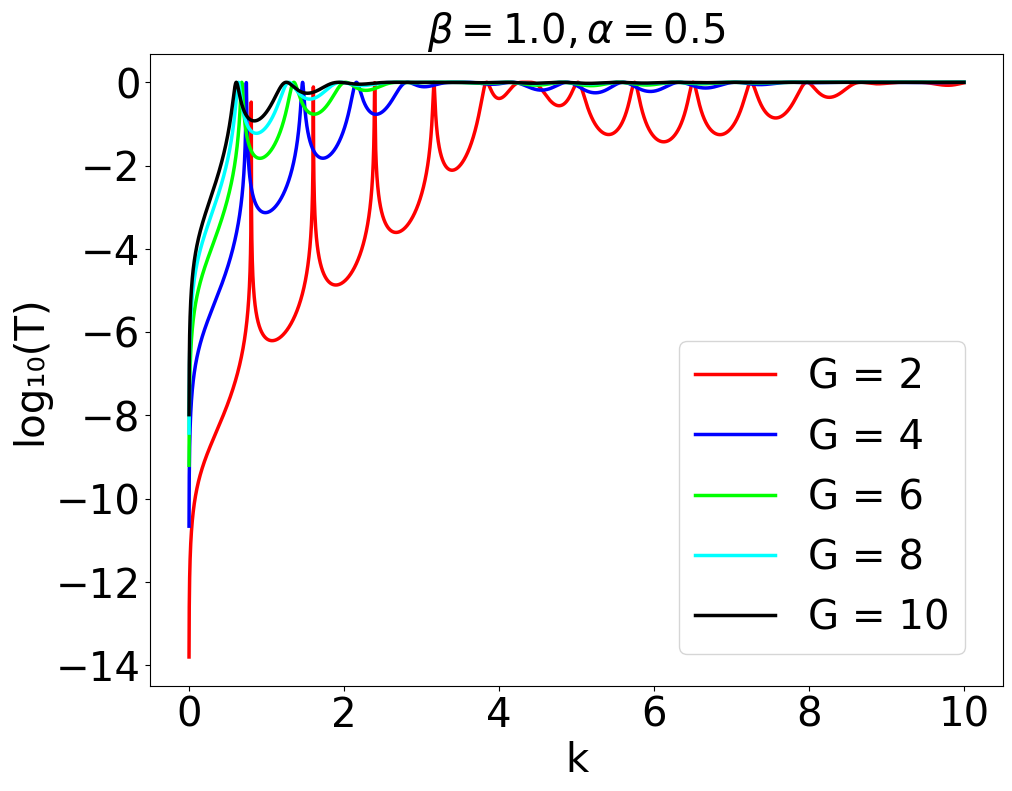} (a)
    \includegraphics[scale=0.2]{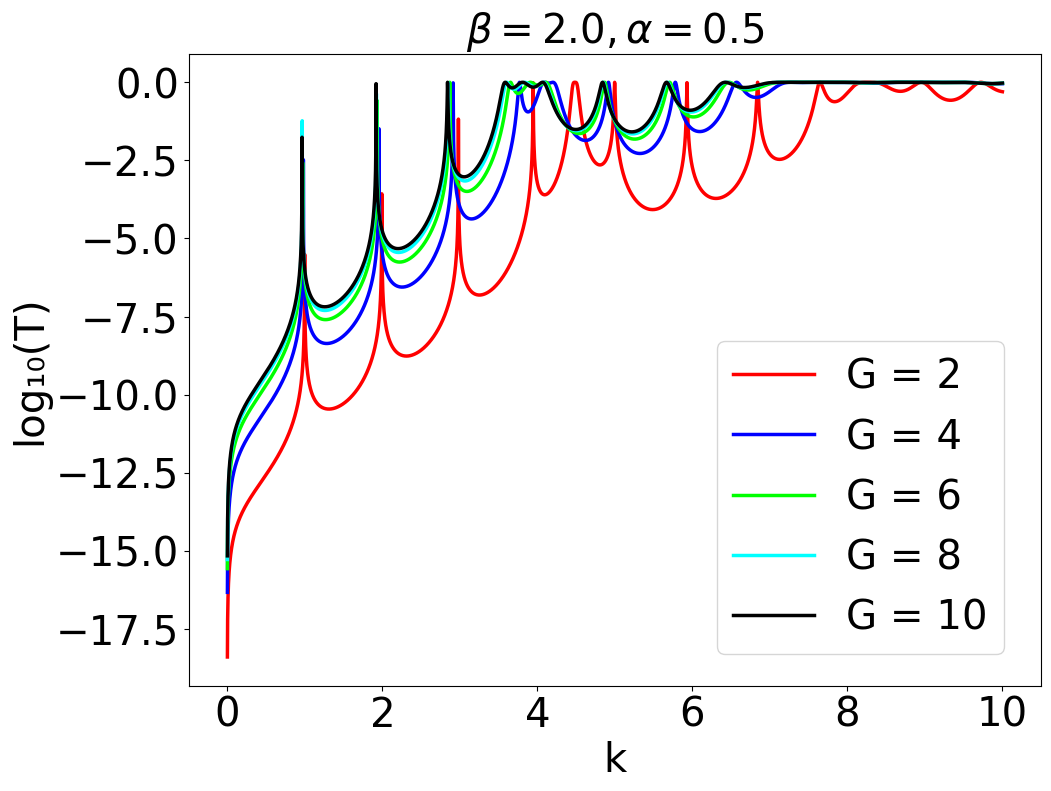} (b)\\
    \includegraphics[scale=0.2]{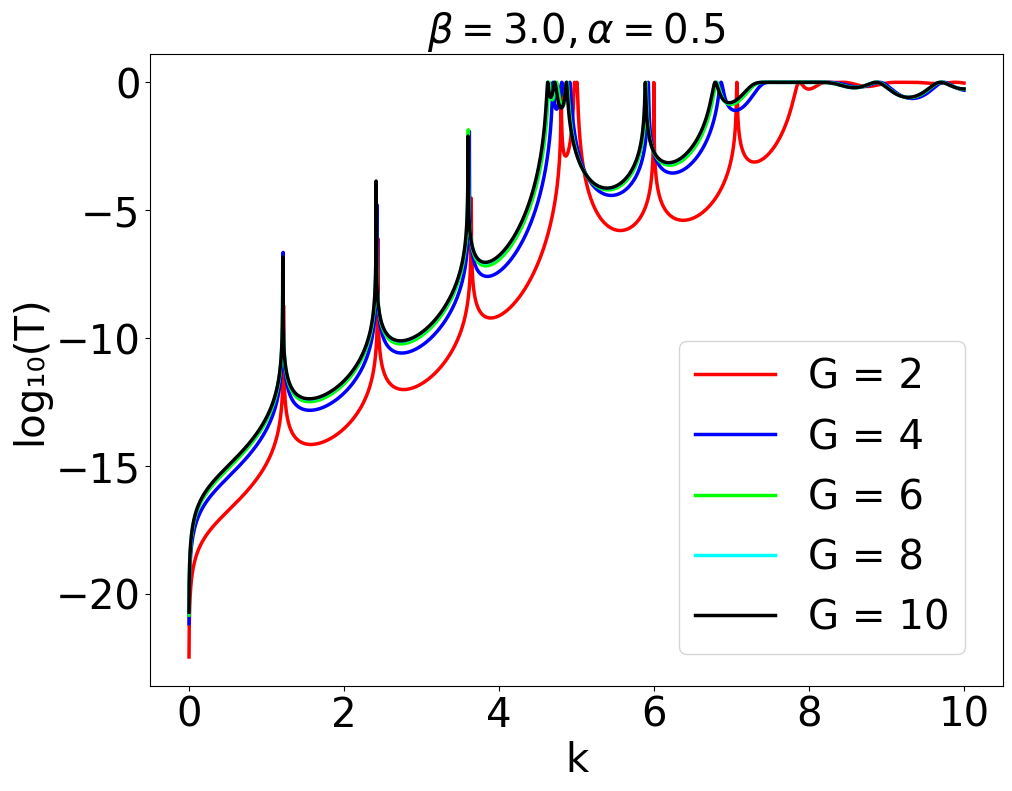} (c)
    \includegraphics[scale=0.2]{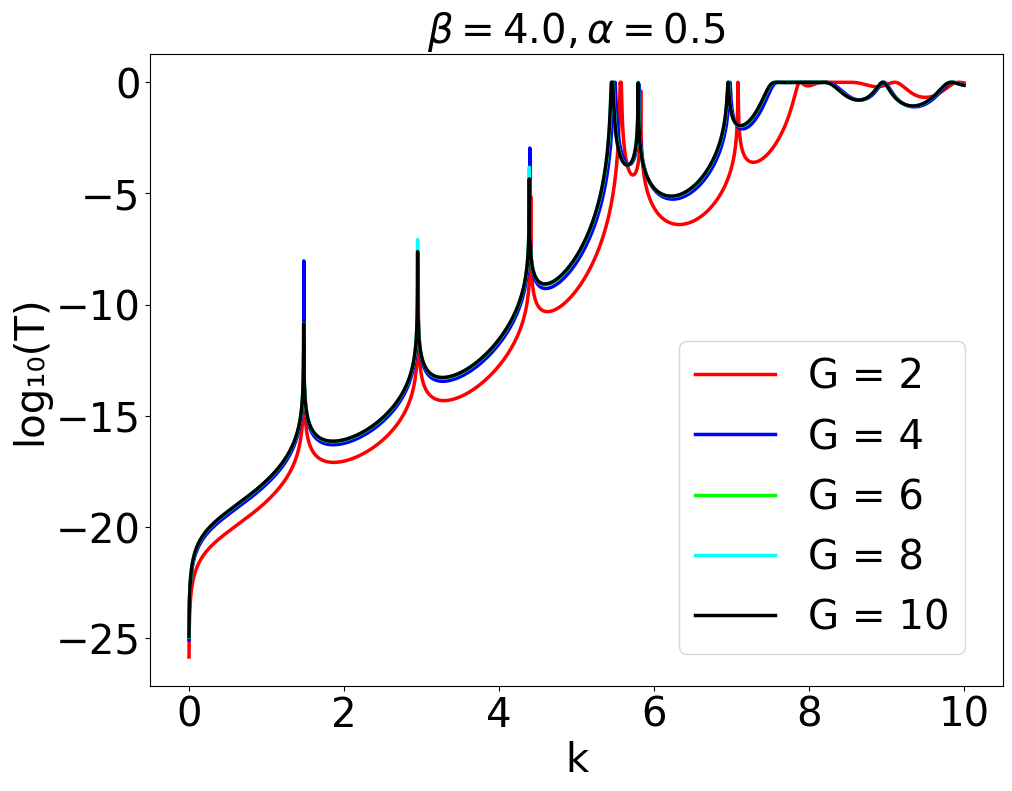} (d)\\
    \includegraphics[scale=0.2]{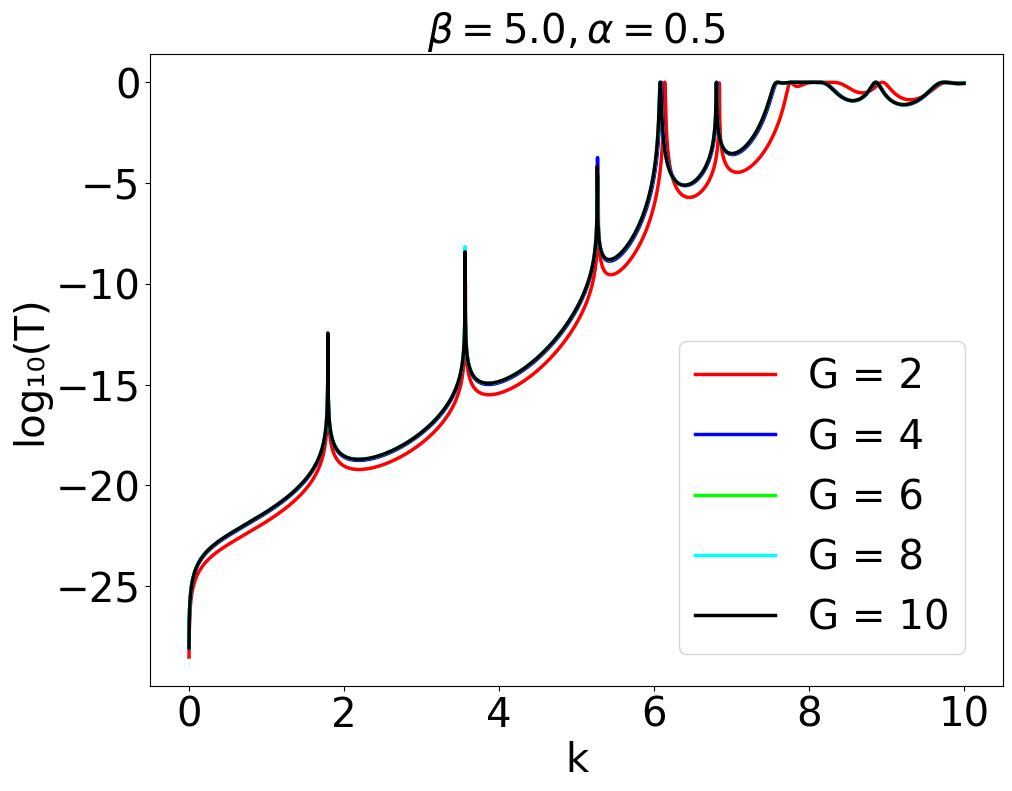} (e)
    \includegraphics[scale=0.2]{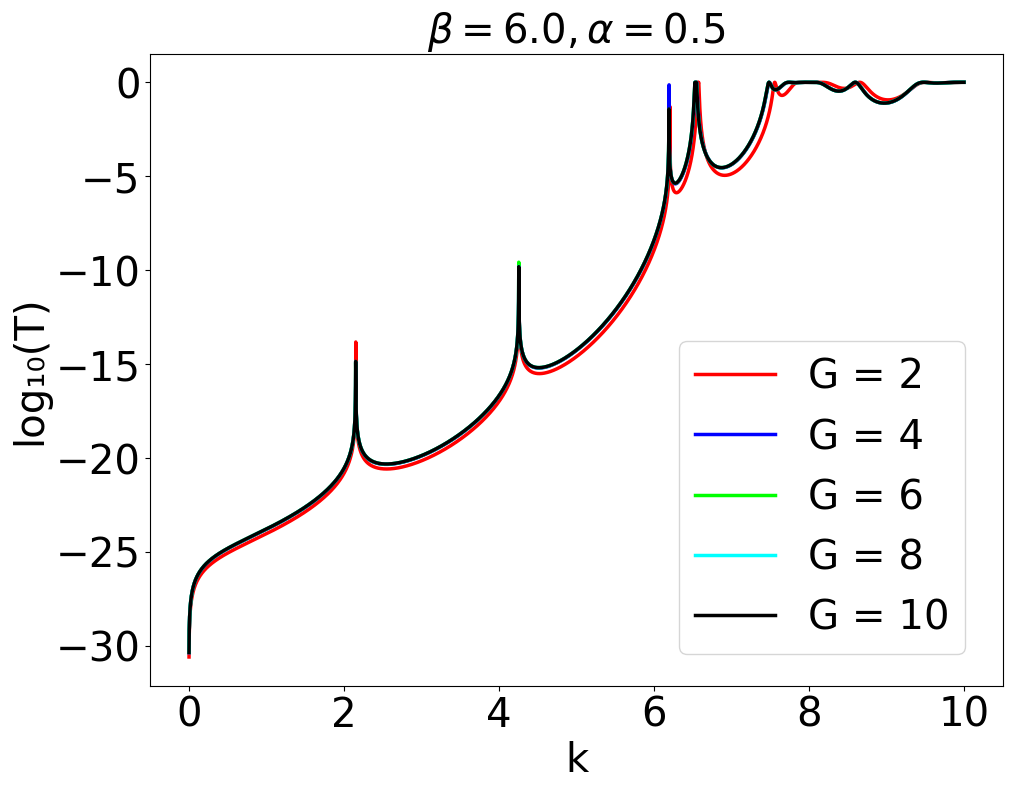} (f)\\
    \includegraphics[scale=0.2]{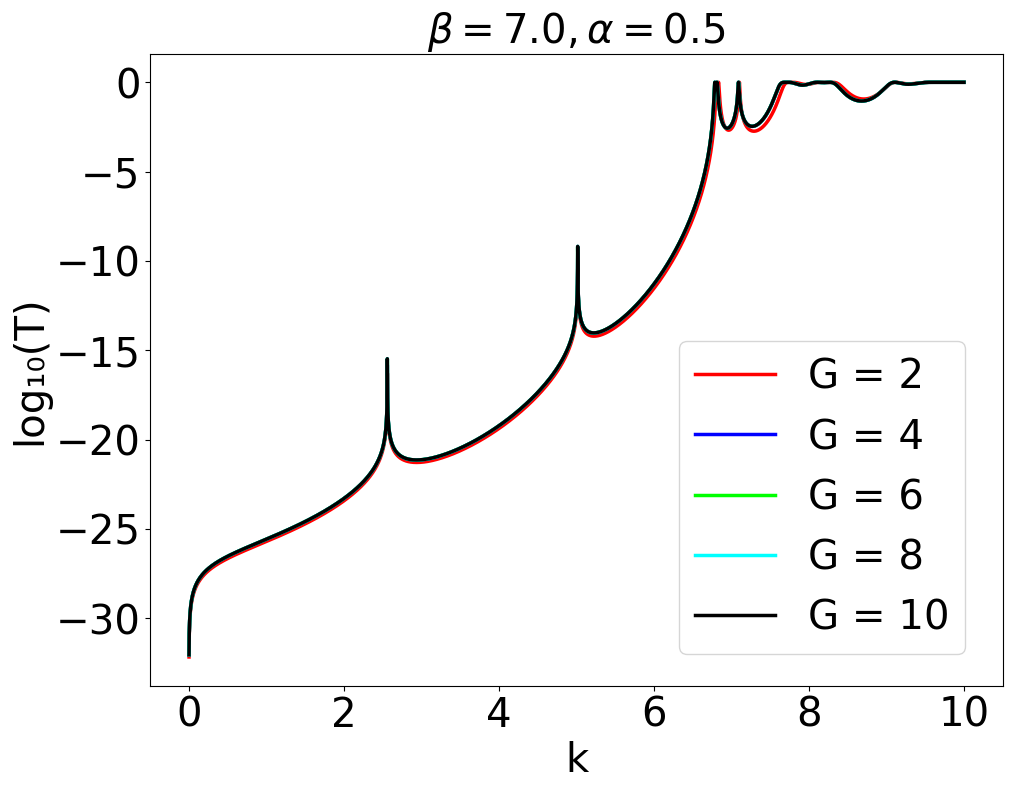} (g)
    \includegraphics[scale=0.2]{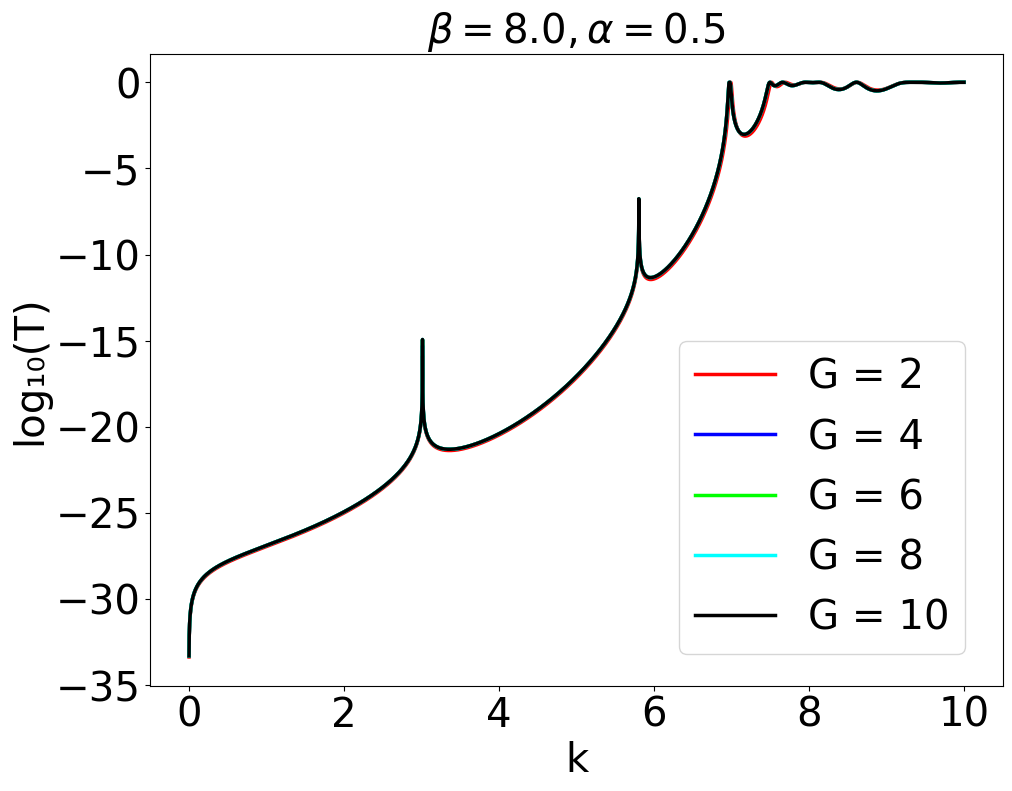} (h)\\
    \includegraphics[scale=0.2]{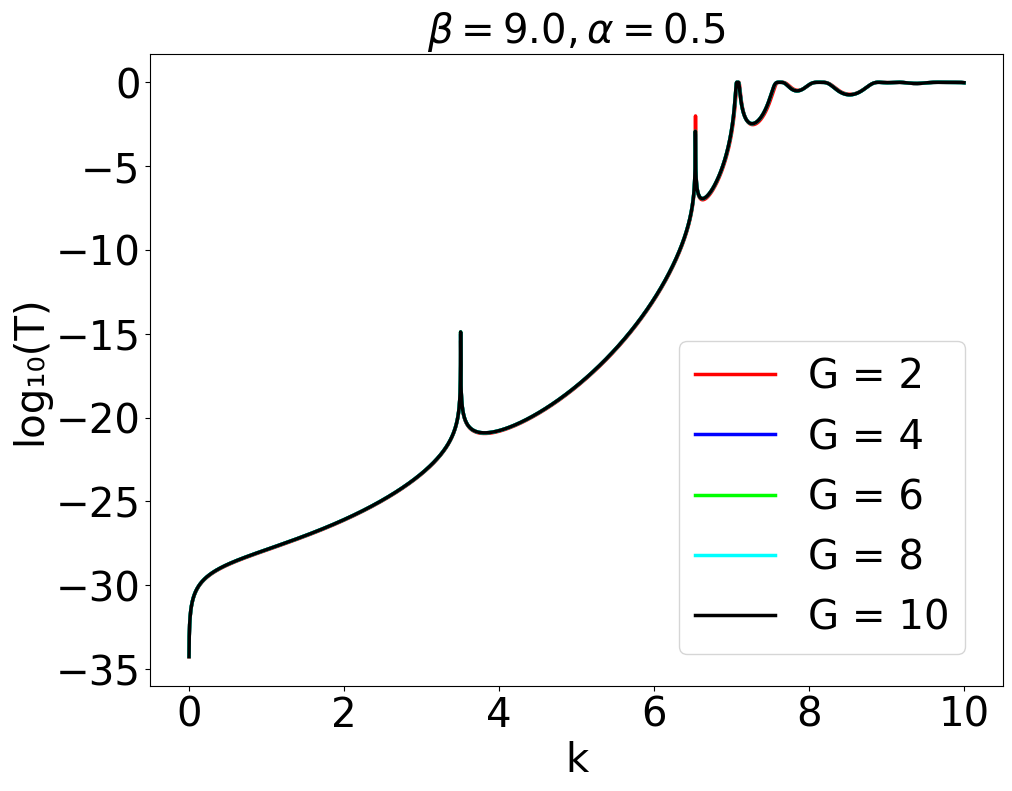} (i)
    \includegraphics[scale=0.2]{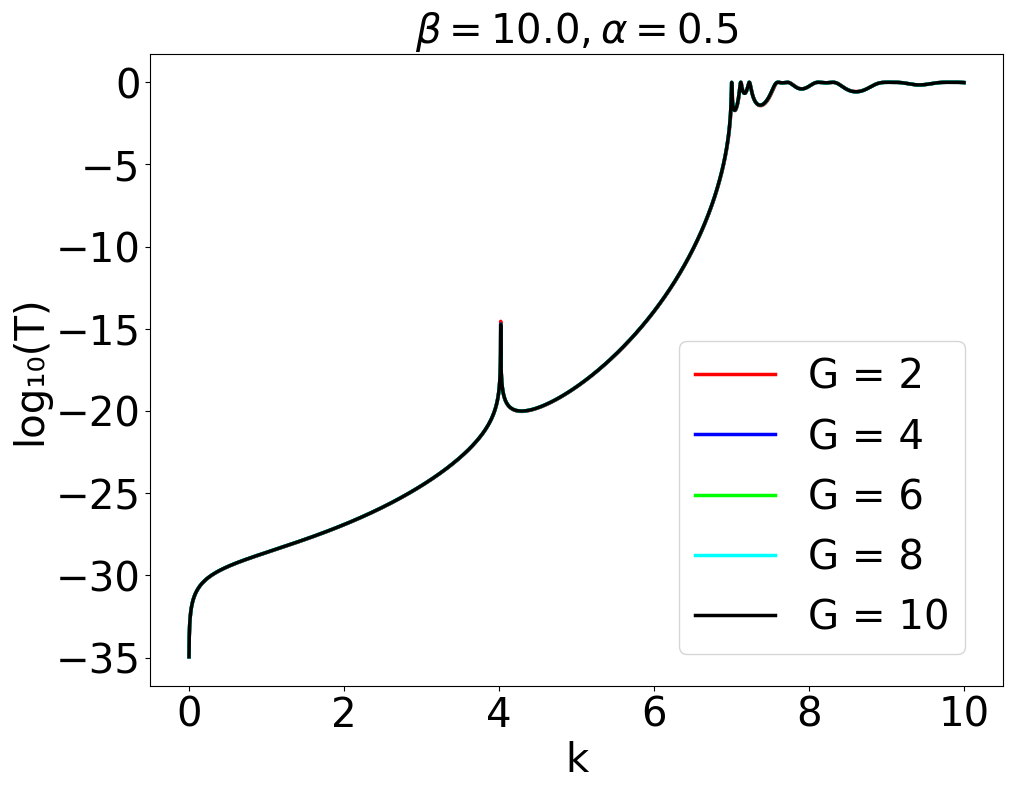} (j)
     \caption{\it{Plots showing the saturation of transmission profile with increasing stage $G$. Here the potential parameters are $L=5$, $V=25$, $\rho= 1.25$ and $\alpha=0.5$. Value of $\beta$ are shown in the figures above. It is seen that for larger $\beta$, saturation occurs for comparatively smaller $G$ values.}}
    \label{logplot}
\end{center}
\end{figure}

The saturation of the transmission profile with $G$ is demonstrated graphically in Fig. \ref{saturationplot} and Fig. \ref{logplot} for different stages $G$ and for different potentials. The potential parameters are $L=5$, $V=25$, $\alpha=0.5$ for both the figures and $\rho=2.5$, 1.25 for Fig. \ref{saturationplot} and Fig. \ref{logplot} respectively. The different plots in the figures are for different values of $\beta$. For a better view of saturation, we have plotted $\log_{10}T(k)$ (y-axis) with $k$. As T lies in the range $0$ to $1$ ($0<T(k)\le1$), therefore $\log_{10}T(k)\le 0$ and is well defined function. From both the figures it is evident that $T(k)$ profile sturates with $G$. Also it is noted that larger value of $\beta$ supports faster saturation with $G$. It is because when $\beta$ is large, more thinner portions from the previous stages are removed for smaller $G$ compared to the case when $\beta$ is small.
\subsection{Extension of the system for $\{\alpha, \beta\} \in \mathbb{R}$}
\label{extension}
\begin{figure}[h!]
\begin{center}
    \includegraphics[scale=0.28]{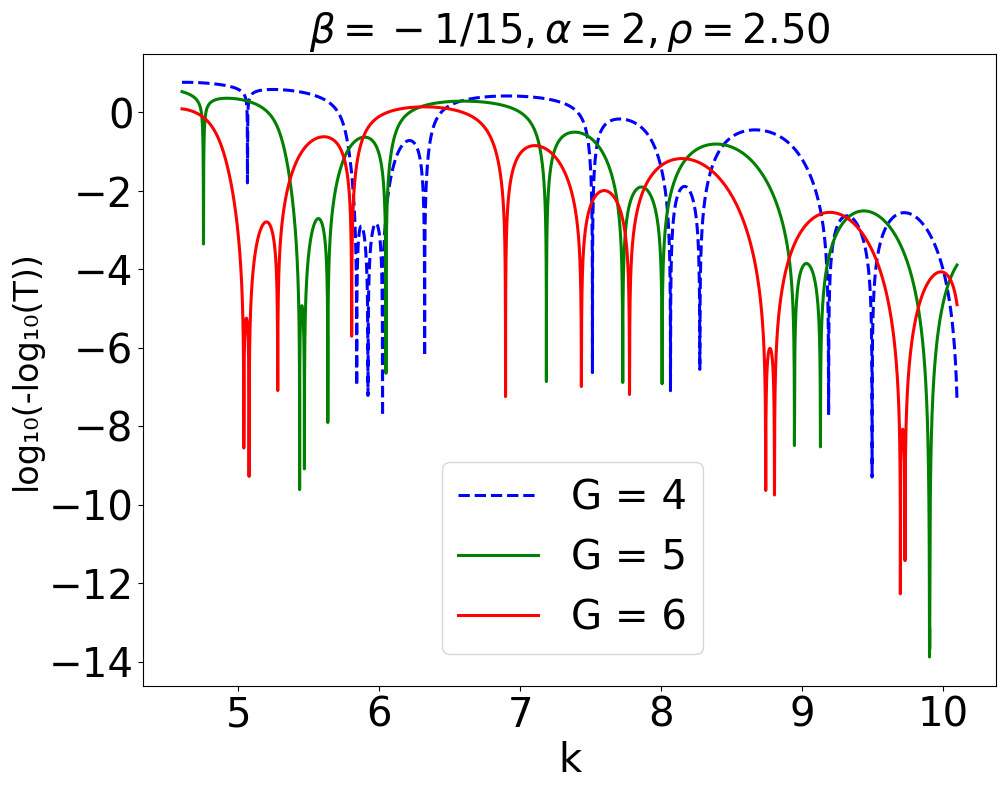} (a)
    \includegraphics[scale=0.28]{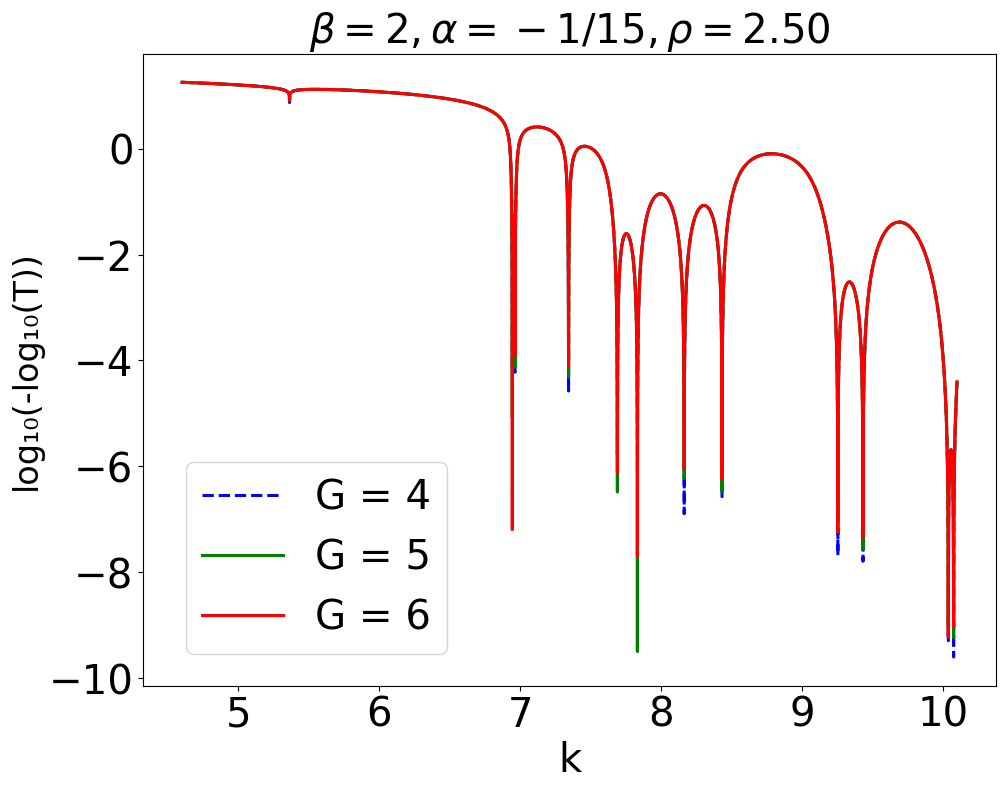} (b)
     \caption{\it{Plot showing the variation of transmission coefficient with k for (a) negative value of $\beta$ and (b) negative value of $\alpha$ for stage $G$ $=$ 4 5 and 6. Here $L=5$, $V=25$, $\alpha=2$ and $\rho=2.5$.}}
    \label{betaalphanegative}
\end{center}
\end{figure}

\begin{figure}[H]
\begin{center}
    \includegraphics[scale=0.26]{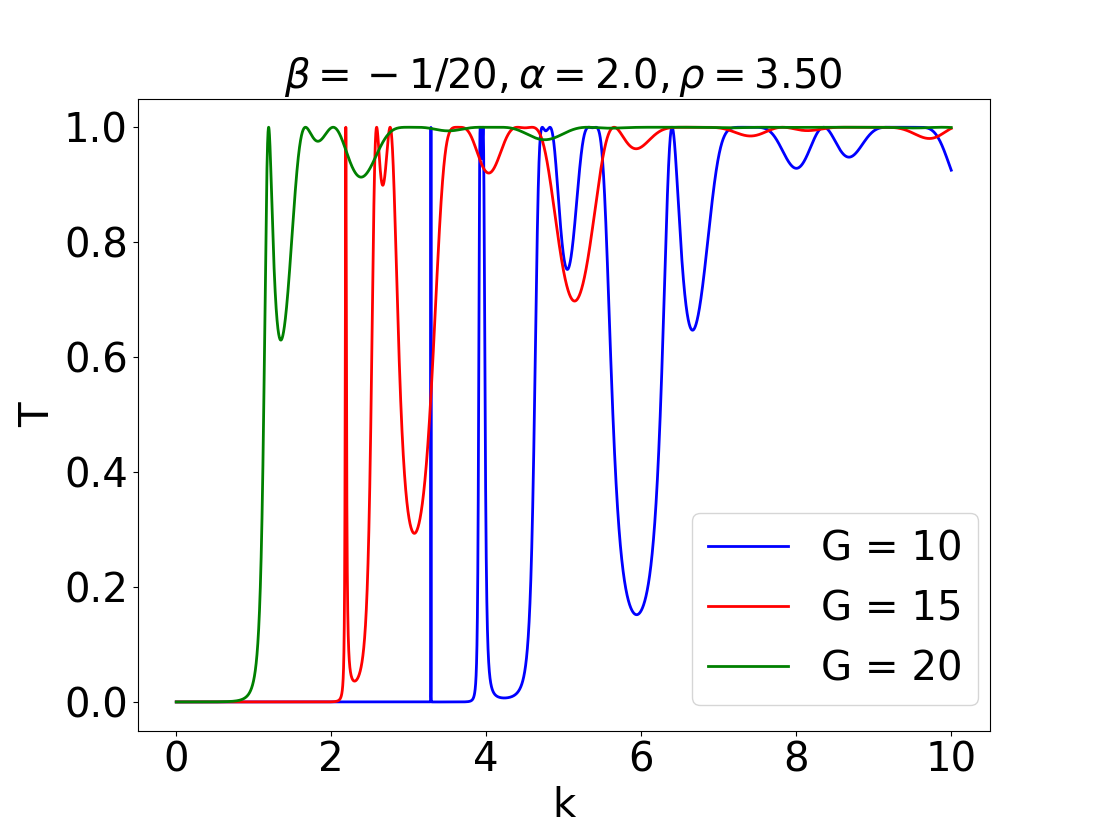} (a)
    \includegraphics[scale=0.26]{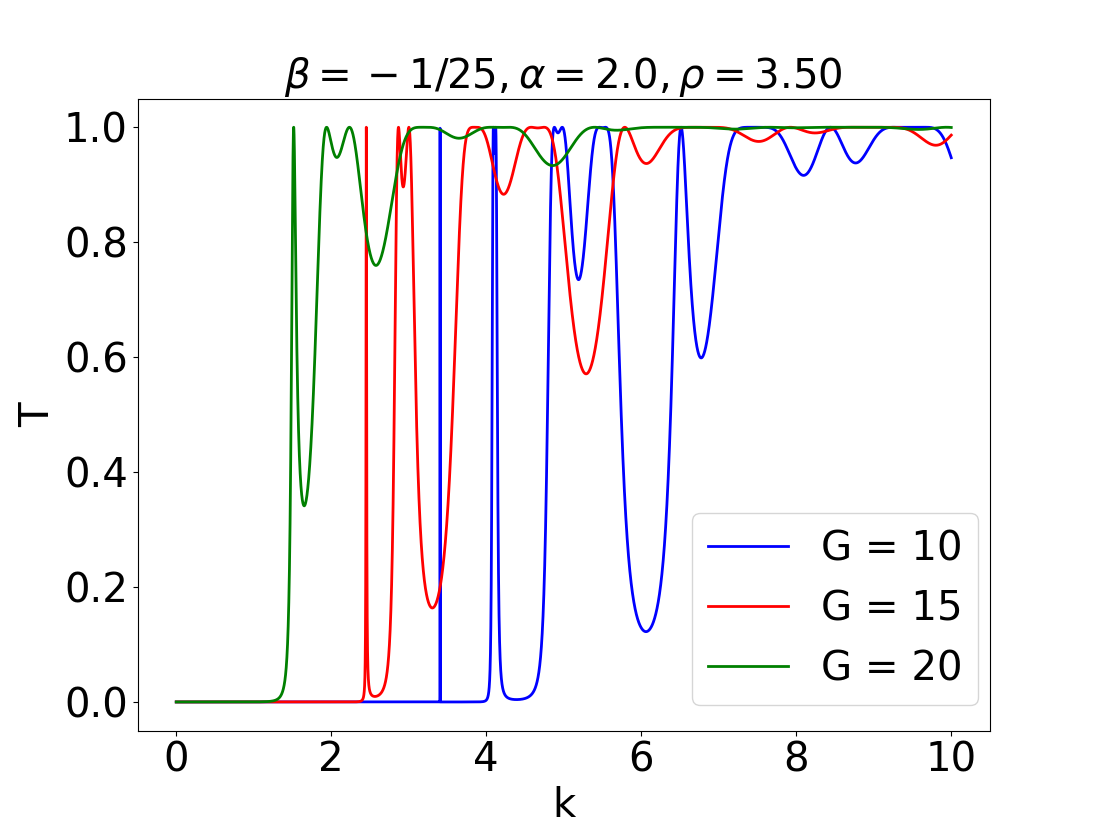} (b)\\
    \includegraphics[scale=0.26]{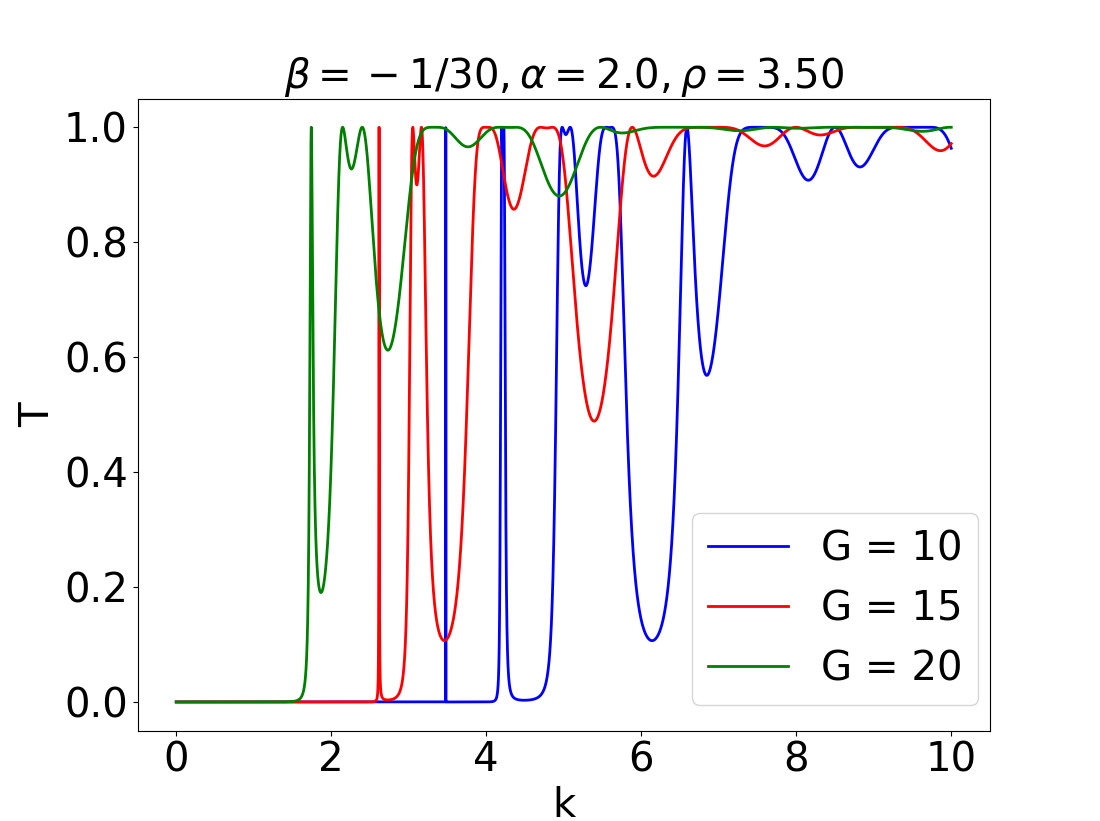} (c)
    \includegraphics[scale=0.26]{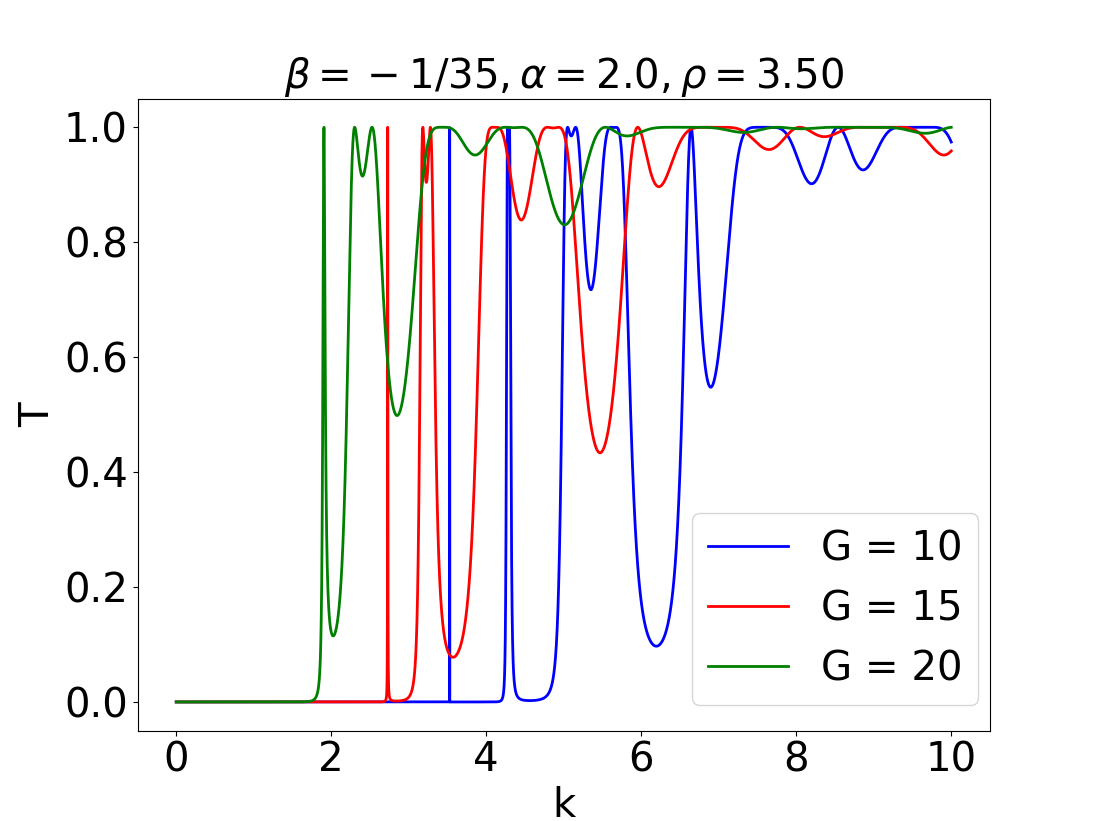} (d)
     \caption{\it{Plot showing the variation of transmission coefficient with k for negative values of $\beta$ for stage $G=$ 10, 15 and 20. Here $L=5$, $V=25$, $\alpha=2$ and $\rho=3.5$.}}
    \label{betanegative}
\end{center}
\end{figure}

\begin{figure}[H]
\begin{center}
    \includegraphics[scale=0.26]{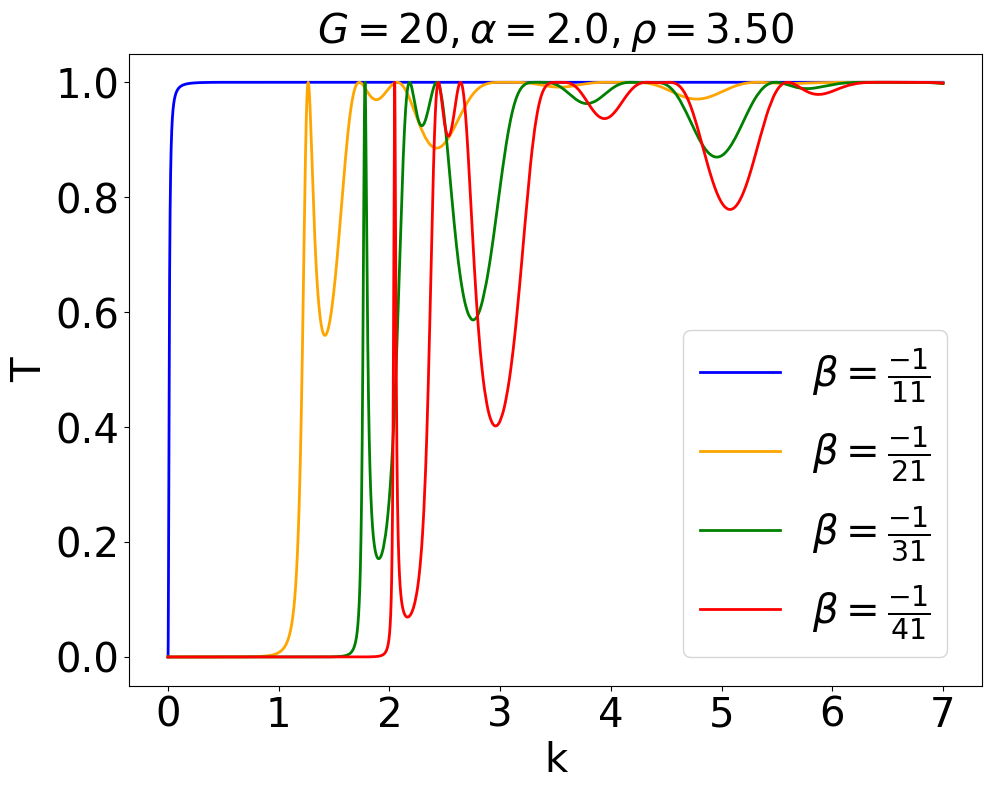}
     \caption{\it{Plot showing the variation of transmission coefficient with k for different negative values of $\beta$ for fixed stage $G=20$, Here $L=5$, $V=25$, $\alpha=2$ and $\rho=3.5$.}}
    \label{betanegative01}
\end{center}
\end{figure}

In this section, we extend the concept of our UCP-$\rho$ potential system for the case when $\alpha$ and $\beta$ can take any real values including negative real numbers. In order to make a well defined  UCP-$\rho$ potential, the following condition should be respected by the geometry of the UCP-$\rho$ potential at a stage $G$
\begin{equation}
    l_{G-1}\left(1-\frac{1}{\rho^{\alpha+\beta G}}\right)>0,
\end{equation}
where $l_{G}$ is the length of the unit potential segment at stage $G$. As $l_{G} >0$, we can express this as 
\begin{equation}
    1-\frac{1}{\rho^{\alpha+\beta G}}>0, 
\end{equation}
which gives $\alpha+\beta G >0$ for $\rho > 1$. This condition shows that for $\rho > 1$ and $\{ \alpha, \beta \} \in \mathbb{R}$, a UCP-$\rho$ potential system is well defined only upto a stage $G$ such that $\alpha+\beta G >0$. For a given combination of $(\alpha$, $\beta$, $\rho)$, it may be possible that the formation of UCP-$\rho$ potential system is not respected for some stage $G$. For example, the combination $(\alpha$, $\beta$, $\rho)$ $=$ $( 2$, $-\frac{1}{10}$, $e)$  will show a localized empty space for stage $G=20$ as it allows the formation of UCP-$\rho$ potential for stage $G<20$ but not for $G\geq20$. Fig. \ref{betaalphanegative} shows the transmission profile for stage $G$ $=$ 4, 5 and $6$ for negative $\beta$  (Fig. \ref{betaalphanegative}a) and negative $\alpha$  (Fig. \ref{betaalphanegative}b). As $0<T\leq 1$, therefore $\log_{10}T(k)\leq 0$ which further implies that $-\log_{10}T(k)\geq 0$. This implies that function $f=\log_{10}(-\log_{10}T(k))$ is a well defined function and we have taken it on y-axis in the given figure to show better resolution of different plots.  In Fig. \ref{betaalphanegative}a, a shift of the transmission profile towards lower energy can be seen as we go to higher stages $G$. Fig. \ref{betaalphanegative}a doesn't show the saturation of transmission profile with $G$ while Fig. \ref{betaalphanegative}b show for the range of $G$ used in the plots. This is due to the fact that for values $(\alpha$, $\beta$, $\rho)$ $=$ $(2$, $-\frac{1}{15}$, $2.5)$, progressively thicker portions from the previous stages are removed as we increase $G$, therefore transmission profiles shown in Fig. \ref{betaalphanegative}a doesn't show saturation with $G$.
However, for Fig. \ref{betaalphanegative}b, where $(\alpha$, $\beta$, $\rho)$ $=$ $(-\frac{1}{15}$, 2, $2.50)$, progressively thinner portions from the previous stages are removed 
as we increase $G$. Therefore the transmission profile show saturation with $G$ as discussed in the previous section. In Fig. \ref{betanegative} we show the transmission resonances for different $\beta <0$ and $G$ $=$ 10, 15 and 20. It is seen that the transmission peaks are more sharper at lower values of $G$ in all these figures. Further in Fig. \ref{betanegative01}, we plot the transmission profile for different $\beta<0$ in the same figure. Transmission resonances appear sharper at lower $\beta$ values in this figure. 
\section{Conclusion and Discussions}
\label{conclusion}
In this paper, we have introduced a new type of potential system that joins the family of general Cantor (GC) and general Smith-Volterra-Cantor (GSVC or SVC-$\rho$) potential systems. We have named this potential as Unified Cantor Potential (UCP) as it unifies two vastly different potential systems i.e., a fractal potential (general Cantor) and a non-fractal potential (SVC-$\rho$). It is an extraordinary fact to note that a fractal system and a non-fractal system can be beautifully combined into one single unified system. Our UCP-$\rho$ system is characterized by scaling parameter $\rho >1$, stage $G$ and two real numbers $\alpha$ and $\beta$. For $\alpha=1$, $\beta=0$, the UCP-$\rho$ system represents general Cantor potential while for $\alpha=0$, $\beta=1$, this system represent SVC-$\rho$ potential.\\
\indent
Further, starting from the concept of super periodic potential (SPP), we have shown that our symmetric UCP-$\rho$ system is a special case of SPP  with unit cell as rectangular potential. Using the concept of SPP, we have derived the analytical expression of transmission coefficient $T_{G}(k)$ for this potential for arbitrary values of $\rho >1$, $\alpha$, $\beta$ and stage $G$ using $q$-Pochhammer symbol. We have provided the density plots of $T_{G}(k)$ for different stages $G$ of UCP-$\rho$ system. Sharp transmission resonances are found from this potential. Very sharp transmission resonances are also noted for SVC-$\rho$ potential \cite{vn_svc} and we believe that the extension of SVC-$\rho$ potential to UCP-$\rho$ would also help in the scope of designing transmission filters of very narrow wavelength range. It is also noted that for $\beta >0$, the tunneling profile with $k$ shows saturation with increasing stage $G$. Further the saturation emerges faster with $G$ for comparatively larger $\beta$ values. This is due to the fact that progressively thinner portions are removed at stage $G$ from previous segments of stage $G-1$.\\
\indent
We also studied the case of the reflection coefficient from UCP-$\rho$ system when the height of the potential changes in a manner that keeps the total area of the potential at each stage $G$ constant. It is shown graphically that the reflection profile rapidly converges with increasing $G$. Further, we have shown analytically that the reflection coefficient will vary as $\frac{1}{k^{2}}$ for large $k$. This is also demonstrated graphically. We have also extended the concept of UCP-$\rho$ system when $\alpha$ and $\beta$ could be negative. In this case, depending upon the values of $\alpha$ and $\beta$, the potential is well defined upto a maximum stage $G$. Also in this case the transmission characteristics are sensitive to the values of $\alpha$ and $\beta$.\\
\indent
We would like to indicate that UCP-$\rho$ system can be further extended towards a broader class of potential that can be realized by removing $\frac{1}{\rho^{a_{0}+ a_{1}G+ a_{2}G^{2}+,..., + a_{n}G^{n}}}$ portion from the middle of each segment from the segments of previous stages. This polynomial type Cantor system will unify a large class of potential systems and can be studied using SPP formalism for transmission characteristics. It will be interesting to study such potential in future work.

\bigskip
\textbf{Acknowledgements}:\\
MU acknowledges the support from OPC Department, IIT Delhi for the encouragement of research activities. BPM acknowledges the support from the research grant under IoE scheme (Number - 6031), BHU, Varanasi. MH acknowledges support from SPO-ISRO HQ for the encouragement of research activities.


\begin{thebibliography}{99}
\bibitem {nordheim} \textit{Electron emission in intense electric fields},  R. H. Fowler and L. Nordheim, Proc. R. Soc. A \textbf{119}, 173 (1928).
\bibitem {gurney} \textit{Quantum mechanics and radioactive disintegration}, R. W. Gurney and E. U. Condon, Phys. Rev. \textbf{33}, 127 (1929).

\bibitem {condon} \textit{Quantum mechanics of collision processes I. Scattering of particles in a definite force field} E. U. Condon and P. M. Morse, Rev. Mod. Phys. \textbf{3}, 43 (1931).

\bibitem {wigner} \textit{Lower limit for the energy derivative of the scattering phase shift}, E. P. Wigner, Phy. Rev. \textbf{98}, 145 (1955).

\bibitem {bohm} \textit{Quantum Theory}, D. Bohm, Prentice-Hall, New York (1951).

\bibitem {book1} \textit{Solvable models in quantum mechnaics}, S. Albeverio, Springer-Verlag New York (1988).

\bibitem {book2} \textit{Quantum theory of tunneling}, M. Razavy, World Scientific (2003).

\bibitem {esaki} \textit{Long journey into tunneling} L. Esaki, Science \textbf{183}, 4130 (1974).

\bibitem {burstein} \textit{Tunneling phenomena in solids}, E. Burstein and S. Lundqvist, New York: Plenum Press, (1969).

\bibitem {giaever} \textit{Electron tunneling and superconductivity}, I. Giaever, Science \textbf{183}, 1253 (1974).

\bibitem {josephson} \textit{The discovery of tunneling supercurrents}, B. D. Josephson, Science \textbf{184}, 527 (1974).

\bibitem {lauhon} \textit{Direct observation of the quantum tunneling of single hydrogen atoms with a scanning tunneling microscope}, L. J. Lauhon and W. Ho, Phys. Rev. Lett. \textbf{85}, 4566 (2000).

\bibitem {angelopoulou1995non} \textit{Non-Hermitian Tunneling of Open Quantum Systems}, P. Angelopoulou, et al., Int. J. of Mod. Phy. B, \textbf{9}, 17(1995). 

\bibitem {hasan2020hartman} \textit{Hartman effect from layered PT-symmetric system}, M. Hasan and B. P. Mandal, Euro. Phys. J. Plus, \textbf{135}, 84 (2020). 
\bibitem {hasan2020role} \textit{Role of PT-symmetry in understanding Hartman effect}, M. Hasan,V. N. Singh and B. P. Mandal, Euro. Phys. J. Plus, \textbf{135}, 640 (2020). 

\bibitem {longhi2022non} \textit{Non-Hermitian Hartman Effect}, S. Longhi, Annalen der Physik, \textbf{534}, 10(2022). 

\bibitem {guo} \textit{Some physical applications of fractional Schr{\"o}dinger equation}, X. Guo and M. Xu, J. Math. Phys. \textbf{47}, 82104 (2006).

\bibitem {oli} \textit{Tunneling in fractional quantum mechanics}, E. C. de Oliveira and J. V. Jr, J. Phys. A: Math. Theor. \textbf{44}, 185303 (2011).

\bibitem {tare} \textit{Transmission through locally periodic potentials in space-fractional quantum mechanics}, J. D. Tare and J. P. H. Esguerra, Physica A \textbf{407}, 43 (2014).

\bibitem {hasan2020tunneling1} \textit{Tunneling time in space fractional quantum mechanics}, M. Hasan and B.P. Mandal, Phys. Letters A, \textbf{382}, 5(2018). 

\bibitem {hasan2020tunneling2} \textit{Tunneling time from locally periodic potential in space fractional quantum mechanics}, M. Hasan and B.P. Mandal, Euro. Phys. J. Plus, \textbf{135}, 1(p127)(2020). 

\bibitem {sobhani} \textit{Scattering in quantum mechanics under quaternionic Dirac delta potential}, H. Sobhani and H. Hassanabadi, Can. J. Phys. \textbf{94}, 15 (2016).

\bibitem {hasan2020new} \textit{New scattering features of quaternionic point interaction in non-Hermitian quantum mechanics}, M. Hasan and B. P. Mandal, JMP \textbf{61}, 3 (2020).
 
\bibitem {hassan01} \textit{Relativistic scattering of fermions in quaternionic quantum mechanics}, H. Hassanabadi, H. Sobhani and A. Banerjee, Eur. Phys. J. C \textbf{77}, 1 (2017).

\bibitem {sobhani01} \textit{New face of Ramsauer--Townsend effect by using a Quaternionic double Dirac potential}, H. Sobhani and H. Hassanabadi, Indian J. Phys. \textbf{91}, 1205 (2017).

\bibitem {de} \textit{A closed formula for the barrier transmission coefficient in quaternionic quantum mechanics}, S. De Leo, G. Ducati, V. Leonardi and K. Pereira, J. Math. Phys. \textbf{51}, 113504 (2010).

\bibitem {davis} \textit{Nonrelativistic quaternionic quantum mechanics in one dimension}, A. J. Davies and B. H. J. McKellar, Phys. Rev. A \textbf{40}, 4029 (1989).

\bibitem {de01} \textit{Analytic plane wave solutions for the quaternionic potential step}, S. De Leo, G. Ducati and T. M. Madureira, J. Math. Phys. \textbf{47}, 082106 (2006).

\bibitem {mandelbrot} \textit{The fractal geometry of nature}, Benoit B. Mandelbrot, San Francisco: W. H. Freeman (1982).

\bibitem {feder} \textit{Fractals}, J. Feder, Plenum Press, New York (1988).

\bibitem {wen} \textit{Subwavelength Photonic Band Gaps from Planar Fractals}, W. Wen, L. Zhou, J. Li, W. Ge, C. T. Chan and P. Sheng, Phys. Rev. Lett. \textbf{89}, 223901 (2002).

\bibitem {shalaev1} \textit{Fractals: Localization of dipole excitations and giant optical polarizabilities}, V. M. Shalaev, R. Botet, D. P. Tsai, J. Kovacs and M. Moskovits, Physica A \textbf{207}, 197-207 (1994).

\bibitem {shalaev2} \textit{Nonlinear Optics of Random Media: Fractal Composites and Metal-Dielectric Films}, V. M. Shalaev, Springer, Berlin (2000).

\bibitem {shalaev3} \textit{Optical Properties of Nanostructured Random Media}, V. M. Shalaev, Springer, Berlin (2002).

\bibitem {takeda} \textit{Localization of Electromagnetic Waves in Three-Dimensional Fractal Cavities}, M. W. Takeda, S. Kirihara, Y. Miyamoto Y, K. Sakoda and K. Honda, Phys. Rev. Lett. \textbf{92}, 093902 (2004).

\bibitem {chuprikov2000} \textit{Electron tunnelling through a self-similar fractal potential on the generalized Cantor set}, N. L. Chuprikov and D. N. Zhabin, J. Phys. A: Math. Gen. \textbf{33}, 4309-4316 (2000).

\bibitem {miyamoto} \textit{Smart Processing Devlopment of Photonic Crystals and Fractals}, Y. Miyamoto,  S. Kirihara, S. Kanehira, M. W. Takeda, K. Honda and K. Sakoda, Int. J. Appl. Ceram. Technol. \textbf{1}, 40-48 (2004).

\bibitem {cantor_graphene} H. Garcia-Cervantes, L. M. Gaggero-Sager, D. S. Diaz-Guerrero, O. Sotolongo-Costa and I. Rodriguez-Vargas, {\em Scientific Reports}, \textbf{7}, 617 (2017).

\bibitem {voss} \textit{Fractals in Nature: From Characterization to Simulation}, R. F. Voss, The Science of Fractal Images, Springer, New York (1988).

\bibitem{hurd} \textit{Resource Letter FR-1: Fractals}, Alan J. Hurd. Am. J. Phys. \textbf{56}, 969 (1988).

\bibitem {cantor_f1} \textit{Scattering on Fractal Measures}, Charles-Antoine Guerin and M. Holschneider, J. Phys. A \textbf{29}, 7651-7667 (1996).

\bibitem {cantor_f2} \textit{Strong Resonance of Light in a Cantor Set}, N. Hatano, J. Phys. Soc. Jpn. \textbf{74}, 3093-3111 (2005).

\bibitem {cantor_f3}\textit{Rigorous solution for electromagnetic waves propagating through pre-Cantor sets}, K. Honda and Y. Otobe, J. Phys. A: Math. Gen. \textbf{39}, L315-L322 (2006).

\bibitem {cantor_f4} \textit{A new type of solution of the Schrodinger equation on a self-similar fractal potential}, N. L. Chuprikov and O. V. Spiridonova, J. Phys. A: Math. Gen. \textbf{39}, L559-L562 (2006).

\bibitem {cantor_f5} \textit{The transfer matrices of the self-similar fractal potentials on the Cantor set},  N. L. Chuprikov, J. Phys. A: Math. Gen. \textbf{33}, 4293-4308 (2008).

\bibitem {cantor_f6} \textit{Wave propagation through Cantor-set media: chaos, scaling, and fractal structures}, K. Esaki, M. Sato and M. Kohmoto, Phys. Rev. E \textbf{79}, 056226 (2009).

\bibitem {cantor_f7} \textit{Scaling laws of reflection coefficients of quantum waves at a Cantor-like potential}, H. Sakaguchi and T. Ogawana, Phy. Rev. E \textbf{95}, 032214 (2017).

\bibitem {cantor_f8} \textit{A transfer matrix method for the analysis of fractal quantum potentials}, Juan A. Monsoriu et al., Eur. J. Phys. \textbf{26}, 603-610 (2005).

\bibitem {cantor_f9} \textit{Scattering on fractal measures}, Guerin, Charles-Antoine and Holschneider, Matthias, J. Phys. A: Math. Gen. \textbf{29}, 7651 (1996).


\bibitem {mh_spp} \textit{Super Periodic Potential}, Mohammad Hasan and Bhabani Prasad Mandal, Annals of Physics \textbf{391}, 240-262 (2018).

\bibitem {vn_svc} \textit{Tunneling through general Smith-Volterra-Cantor fractal potential}, Vibhav N. Singh, Mohammad Umar, Mohammad Hasan and Bhabani Prasad Mandal, J. Math. Phys. \textbf{64}, 032101 (2023).

\bibitem {vn_sfqm} \textit{Quantum tunneling from family of Cantor potentials in fractional quantum mechanics}, Vibhav N. Singh, Mohammad Umar, Mohammad Hasan and Bhabani Prasad Mandal, Annals of Physics \textbf{450}, 169236 (2023).

\bibitem {griffiths} \textit{Waves in locally periodic media}, Griffiths, David J and Steinke, Carl A, American Journal of Physics \textbf{69}, 137-154 (2001).

\bibitem {abramowitz1964} \textit{Handbook of mathematical functions with formulas graphs, and mathematical tables}, Abramowitz, Milton and Stegun, Irene A, US Government printing office \textbf{55}, (1964).


\end{thebibliography}
\end{document}